\documentclass[traditabstract]{aa}  
\usepackage{graphbox,graphicx}
\usepackage[online]{threeparttablex}
\usepackage[dvipsnames]{xcolor}
\usepackage{graphicx,times,txfonts}
\usepackage{natbib}
\usepackage{color}
\usepackage{multirow}
\usepackage{longtable}
\usepackage{todonotes}
\usepackage{appendix}
\usepackage{changepage}
\usepackage{array}
\usepackage{verbatim}

 \usepackage{booktabs}
\usepackage{subcaption}
\usepackage{mwe}
\usepackage{lscape} 
\usepackage{multirow}
\usepackage{tablefootnote}
\usepackage{siunitx}
\usepackage{booktabs}
\usepackage{color, colortbl}
\usepackage{hyperref}
\hypersetup{
    colorlinks=true,
    linkcolor=blue,
    citecolor=blue,
    filecolor=black,      
    urlcolor=black,
    }
\usepackage{varwidth}
\usepackage{placeins}
\newsavebox\tmpbox
\newcommand{\planck}{{\it Planck}}
\newcommand{\xmm}{{\it XMM-Newton}}

\begin{document} 

\title{CHEX-MATE: Morphological analysis of the sample}

   \author{M. G. Campitiello\inst{1,2} \fnmsep\thanks{\email{giulia.campitiello@inaf.it}}
          \and
          S. Ettori\inst{1,3}
          \and
          L. Lovisari\inst{1,4} 
          \and
          I. Bartalucci\inst{5}
          \and 
          D. Eckert\inst{6}
          \and
          E. Rasia\inst{7,8}
         \and
          M. Rossetti\inst{5}
          \and
          F. Gastaldello\inst{5}
          \and
          G.W. Pratt\inst{9}
          \and
          B. Maughan\inst{10}
          \and
          E. Pointecouteau\inst{11}
          \and
          M. Sereno\inst{1,3}
          \and
          V. Biffi\inst{12,13}
          \and
          S. Borgani\inst{7,12,14,15}
          \and
          F. De Luca\inst{16}
          \and
          M. De Petris\inst{17}
          \and
          M. Gaspari\inst{1,18}
          \and 
          S. Ghizzardi\inst{5}
          \and
          P. Mazzotta\inst{16}
          \and
          S. Molendi\inst{5}
                    }

   \institute{INAF, Osservatorio di Astrofisica e Scienza dello Spazio, via Piero Gobetti 93/3, 40129 Bologna, Italy
         \and
         Dipartimento di Fisica e Astronomia, Università di Bologna, Via Gobetti 92/3, 40121, Bologna, Italy
         \and
         INFN, Sezione di Bologna, viale Berti Pichat 6/2, 40127 Bologna, Italy
         \and 
        Center for Astrophysics $|$ Harvard $\&$ Smithsonian, 60 Garden Street, Cambridge, MA 02138, USA
        \and 
        INAF – Istituto di Astrofisica Spaziale e Fisica Cosmica di Milano, Via A. Corti 12, 20133 Milano, Italy
        \and
         Department of Astronomy, University of Geneva, ch. d’Ecogia 16, 1290 Versoix, Switzerland
         \and
         Dipartimento di Fisica, Sezione di Astronomia, Università di Trieste, via Tiepolo 11, 34143 Trieste, Italy
         \and
         Institute of Fundamental Physics of the Universe, via Beirut 2, 34151 Grignano, Trieste, Italy
         \and
         AIM, CEA, CNRS, Universit\'e Paris-Saclay, Universit\'e Paris Diderot, Sorbonne Paris Cit\'e, F-91191 Gif-sur-Yvette, France
         \and
         H. H. Wills Physics Laboratory, University of Bristol, Tyndall Avenue, Bristol BS8 1TL, UK
         \and
         IRAP, Université de Toulouse, CNRS, CNES, UPS, 9 av du colonel
         Roche, BP 44346, 31028 Toulouse Cedex 4, France
         \and
         IFPU - Institute for Fundamental Physics of the Universe, Via Beirut 2, 34014 Trieste, Italy
         \and
         University Observatory Munich, Scheinerstr. 1, D-81679 Munich, Germany
        \and
         INAF - Osservatorio Astronomico di Trieste, via G. B. Tiepolo 11, I-34143 Trieste, Italy
         \and
         INFN–Sezione di Trieste, Trieste, Italy
         \and
         Università degli studi di Roma `Tor Vergata', Via della ricerca scientifica, 1, 00133, Roma, Italy
         \and
         Dipartimento di Fisica, Sapienza Università di Roma, Piazzale Aldo Moro 5, I-00185, Roma, Italy
         \and
         Department of Astrophysical Sciences, Princeton University, 4 Ivy Lane, Princeton, NJ 08544-1001, USA
         }

   \date{Received 3 March 2022 / Accepted 20 May 2022}

 
  \abstract
{A classification of the galaxy clusters dynamical state is crucial when dealing with large samples. The identification of the most relaxed and most disturbed objects is necessary for both cosmological analysis, focused on spherical and virialized systems, and astrophysical studies, centred around all those micro-physical processes that take place in disturbed clusters (such as particle acceleration or turbulence). To the most powerful tools for the identification of the dynamical state of clusters belongs the analysis of their intracluster medium (ICM) distribution. In this work, we performed an analysis of the X-ray morphology of the 118 CHEX-MATE (Cluster HEritage project with XMM-Newton – Mass Assembly and Thermodynamics at the Endpoint of structure formation) clusters, with the aim to provide a classification of their dynamical state. To investigate the link between the X-ray appearance and the dynamical state, we considered four morphological parameters: the surface brightness concentration, the centroid shift, and the second- and third-order power ratios. These indicators result to be: strongly correlated with each other, powerful in identifying the disturbed and relaxed population, characterised by a unimodal distribution and not strongly influenced by systematic uncertainties. In order to obtain a continuous classification of the CHEX-MATE objects, we combined these four parameters in a single quantity, M, which represents the grade of relaxation of a system. On the basis of the M value, we identified the most extreme systems of the sample, finding 15 very relaxed  and 27 very disturbed galaxy clusters. From a comparison with previous analysis on X-ray selected samples, we confirmed that the Sunyaev–Zeldovich (SZ) clusters tend to be more disturbed. Finally, by applying our analysis on a
simulated sample, we found a general agreement between the observed and simulated results, with the only exception of the concentration. This latter behaviour, is partially related to the presence of particles with high smoothed-particle hydrodynamics density in the central regions of the simulated clusters due to the action of the idealised isotropic thermal Active Galactic Nuclei (AGN) feedback.}

   \maketitle
%

\section{Introduction}

Clusters of galaxies represent a common ground between astrophysics and cosmology. On the one hand, they are the most massive and virialized systems in the Universe and provide a unique opportunity to study processes related to structure formation on both large and small scales, such as cluster and galaxy scales, respectively. On the other hand, the description of their abundance and spatial distribution allows us to derive relevant information on the underlying cosmology, the gravitational processes, and the initial conditions that characterised our Universe. These two areas of research focus on objects that throughout their life experience very different dynamical states, from being very relaxed to very disturbed due, for instance, a major merger and other astrophysical processes (such as turbulence, feedback or feeding flows). Relaxed systems are particularly suitable to derive the cluster total mass, which is, together with the redshift, the most important cluster property used in cosmology. Given the absence of sign of mergers and turbulence, the mass of relaxed systems is derived assuming that both the Intracluster Medium (ICM) and the galaxies are in hydrostatic equilibrium (HE) within the binding cluster potential \citep[see, e.g.,][]{Ettori2013,Pratt2019}. The X-ray mass estimation turns out to be close to gravitational lensing mass \citep[e.g.,][]{Meneghetti2010, Rasia2012} and consequently it is considered robust. Disturbed systems, do not satisfy the HE assumption and their mass estimations are characterised by larger uncertainties. However, astrophysical studies mainly focus on these systems, since phenomena such us turbulence are more prominent. For example, since a correlation between the X-ray emission and the presence of giant radio halos has been observed in merging systems, disturbed clusters are considered the perfect laboratories for the study of particle acceleration mechanisms resulting in non-thermal radio emission \citep[e.g.,][]{cassano2010,Mann2012}. 

The identification of the most relaxed and disturbed systems is essential also for the understanding of the absolute scatter that characterises the scaling relations. This topic is one of the most important open issues in the study of clusters, since it is linked to the constraints to use for cosmological models \citep[e.g.,][]{Lima2005}. It was found that relaxed and disturbed objects lie on different regions of the scaling relations and that their dynamical state provides the major contribution to the scatter about the relations \citep[e.g.,][]{Pratt2009,2020ApJ...892..102L}. For example, \cite{Fabian1994} noticed for the first time that the offset of a cluster from the mean relation is linked to the presence of a cool core, a typical feature of relaxed systems, while \cite{andrade2012}  showed that disturbed systems can be used in the scaling relations when the level of substructures is known and parametrised, so that their positions in the mass-observable planes can be corrected \citep{Ventimiglia2008}. 

Finally, the dynamical state is crucial also in the interpretation of survey data, because of its impact on the selection function. Merger events could influence the observable used to detect clusters, affecting their identification and selection, and increasing (or decreasing) the number of objects observed with respect to what is expected from theoretical mass function. 
For example, by comparing \planck\ Sunyaev-Zeldovich effect (SZE, \citealt{sz72}) clusters with X-ray selected samples, it was found that the latter ones tend to detect mainly centrally peaked and more relaxed clusters. This behaviour is related to the different dependence of
the Sunyaev–Zeldovich (SZ) signal and X-ray emission on the gas density. Since the X-ray emission scales with the square of the gas density, X-ray surveys tend to preferentially detect centrally peaked, more relaxed galaxy clusters \citep{Eckert2011}, which result to be  more luminous at a given mass. On the contrary, the SZ signal is less sensitive to the central gas density and simulations have shown that SZ-surveys are not strongly influenced by the dynamical state of the clusters \citep{Motl2005}. Hence, SZ samples are expected to provide a clean reconstruction of the underlying cluster population. This is a key property for those statistical cluster studies, that aim to constrain cosmological models or to probe the physics of structure formation.

In this context, it is clear that the dynamical classification plays an important role when dealing with large samples of galaxy clusters, since it allows either to identify the most suitable set of systems to consider in the analysis or to estimate any systematic effect introduced by the relative fraction of relaxed and disturbed systems. However, obtaining an accurate characterisation of the dynamical state of galaxy clusters is very challenging because multi-wavelength information are required and they are available only for a few objects. To overcome this limitation it is possible to resort to the analysis of the distribution of the X-ray emission of galaxy clusters: all those processes that can alter the dynamical state of clusters, like mergers, are indeed expected to leave traces in the ICM distribution, and thus in the X-ray images. Therefore, the identification of a proper method for the characterisation of the X-ray morphology, has drawn the attention of the X-ray community over the past thirty years favouring the development of many procedures. Initially, images were inspected by eye to detect and characterise substructures. For example, \cite{Jones1992} distinguished clusters in: ‘single’,‘double’, ‘primary with small secondary’, ‘complex’, ‘elliptical’ (according to the X-ray contours), ‘off-centre’ (either presenting a difference between the centres in optical and X-ray or showing an X-ray tail extended only in one sector off the X-ray peak), and ‘galaxy’ (when the main contribution to the X-ray emission is provided by the central galaxy). However, a classification based only on the presence of substructures may not include all those systems that have not recently interacted with merging massive systems, but despite this, show traces of previous interactions in form of either a strongly elliptical shape \citep{Buote1996,Pinkney1996,Plionis2002}, or a variation of their X-ray centroid \citep{Mohr1995}. Furthermore, the visual classification turns out to be both subjective, since different researcher may provide a different classification for individual clusters, and very time-consuming in case of large samples of clusters like those expected from future surveys. It then became necessary to define more robust indicators able to objectively quantify even small deviations from a perfectly regular and spherically-symmetric emission. Among the commonly used parameters, we point out in particular the axial ratios \citep{Mohr1993}, the centroid shift \citep{Mohr1995,OHara2006}, the power ratios \citep{buote1995,Buote1996}, the light concentration \citep{Santos2008}, and two parameters arising from morphological analysis of galaxies, i.e., the asymmetry and the smoothness \citep{Lotz2004}. To make the classification more effective and least affected by bias and projection effects, these and other parameters were also used in combination \citep[e.g.,][]{Rasia2013,mantz2015,andrade2017,Lovisari2017,McDonald2017,nurgaliev2017,Ghirardini2021}. 

\begin{figure*}[t!]
    \centering
    \includegraphics[scale=0.5]{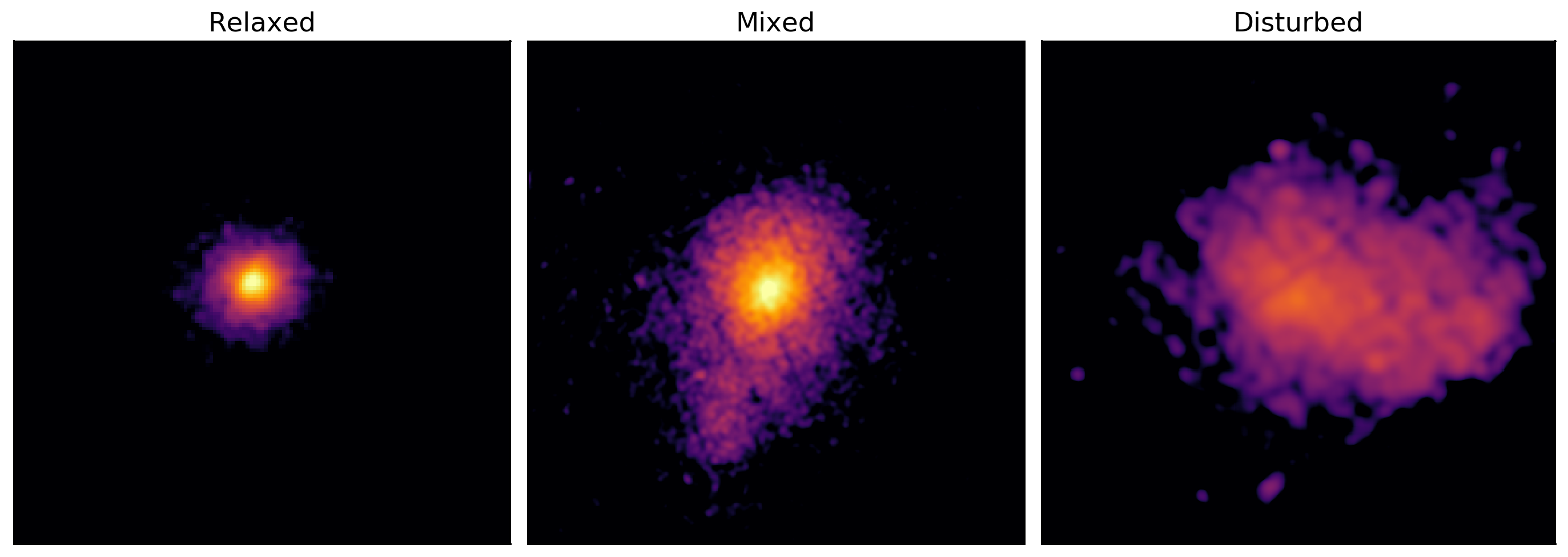}
    \caption{Example of relaxed (left), mixed (centre) and disturbed (right) systems identified by the visual classification. The horizontal and vertical dimensions of the boxes are equal to 2$R_{500}$.}
    \label{fig:example_rmd}
\end{figure*}

In recent years, this approach has been joined with morphological analysis of simulated samples \citep{Rasia2013,Barnes2018}. In the case of hydrodynamical simulations, the cluster state is typically parametrised though a few indicators that are evaluated in 3D. These consider and quantify various conditions of a dynamically active object, such as the presence of well identified substructures (indicating recent mergers); the displacement between the barycentre and the total density peak or the minimum of the potential well (indicating either past merger activities or minor merger wit a small impact parameter), the ratio between the kinetic or thermal energy and the gravitational energy. These parameters allow us to obtain a prior knowledge of the dynamical state of the considered systems \citep[e.g.,][]{Cialone2018,Cui2018,Capalbo2021,DeLuca2021}. For this reason, simulations can be used to calibrate and check the robustness of the morphological parameters \citep[e.g.,][]{Weissmann2013}. From the comparison between observations and simulations it was found that, when used on single clusters, the morphological parameters should be used with extreme caution because they could be affected by substantial uncertainties \citep{Boehringer2010}. \cite{Jeltema2008}, for example, find that less than one half of clusters which are classified as relaxed by the power ratios are truly relaxed, and 4\% - 10\% of these relaxed clusters have very disturbed morphologies when viewed from other orientations. 
On the other hand, not only observations are limited by projection, resolution and background, but also simulations are still unable to capture the complex multi-scale physics of the core, especially with regards to the AGN physics implementation \citep[][for a review]{Gaspari2020}. 

In this paper, we perform a morphological analysis of the X-ray images obtained from \xmm\ for the 118 clusters that constitute the CHEX-MATE (Cluster HEritage project with XMM-Newton – Mass Assembly and Thermodynamics at the Endpoint of structure formation) sample. Aim of the CHEX-MATE project is to set the stage of future X-rays missions, by providing both an overview of the statistical properties of the underlying cluster population and an improvement of the analysis techniques developed up to date. The morphological analysis itself reflects these two goals. The complete and homogeneous X-ray exposures of the CHEX-MATE objects, allow to derive for the first time a uniform characterisation of the X-ray morphology of the cluster underlying population, providing to the entire community an overview of the dynamical state of the CHEX-MATE clusters, which will be useful for the identification of the proper set of systems to use in specific analyses. Furthermore, our analysis aims to  check the techniques developed until now, by testing the efficiency of the morphological parameters.

The paper is structured as follows. In Sect. 2, we present the sample and the dataset that we will analyse. In Sect. 3, we describe the results of a preliminary visual classification of the CHEX-MATE sample, which will be used as reference for the interpretation of the results. In Sect. 4, we describe the set of morphological parameters that we will estimate from the X-ray images. In Sect. 5, we report the results of the morphological analysis carried out on the X-ray observations. In Sect. 6, we investigate the robustness of the morphological parameters, by assessing the systematic uncertainties that affect their measurements. In Sect. 7 we combined them to obtain a unique indicator of the grade of relaxation of clusters. In Sect. 8, we discuss our results and we draw our conclusions in Sect. 9. Throughout the paper, if not otherwise stated, we assume a flat $\Lambda$CDM cosmology with $\Omega_m$ = 0.3, $\Omega_{\Lambda}$ = 0.7 and $H_0$ = 70 km s$^{-1}$Mpc$^{-1}$. The variables $M_{\Delta}$ and $R_{\Delta}$ are the total mass and radius corresponding to a total density contrast $\Delta\rho_c(z)$, where $\rho_c(z)$ is the critical density of the Universe at the cluster redshift (for example, $M_{500}$ = (4$\pi$/3)500$\rho_c(z)R^3_{500}$).



\section{Dataset}\label{dataset}

\subsection{The CHEX-MATE sample}\label{CHEX_MATE_sample}
The CHEX-MATE programme\footnote{http://xmm-heritage.oas.inaf.it/} is described in detail in \cite{CHEXMATE}.
It is a 3 mega-second Multi-Year \xmm\ Heritage Programme to obtain X-ray observations of a minimally-biased, signal-to-noise-limited sample of 118 galaxy clusters detected by \planck\ through the SZE.
The project has been developed to provide an accurate vision of the statistical properties of the underlying population of clusters, to measure how the gas properties are shaped by collapse into the dark matter halo, to uncover the origin of non-gravitational heating, and to resolve the major uncertainties in mass determination that limit the use of clusters for cosmological parameter estimation. To achieve these aims, a sample of 118 Planck clusters \citep{Planck8,Planck27,Planck29}, populating two different sub-samples, was selected through their SZE signal ($S/N$> 6.5) accordingly to the following criteria:

\begin{itemize}
    \item the Tier 1, consisting of 61 objects located at low redshift in the Northern sky (0.05 < $z$ < 0.2 and DEC > 0, with 2 $\times$ 10$^{14}$ $M_{\odot}<$ $M_{500}<$ 9 $\times$ 10$^{14}$ $M_{\odot}$) and providing an unbiased view of the population at the most recent time;
    \item the Tier 2, including the most massive systems to have formed thus far in the history of the Universe ($z$ < 0.6 with $M_{500}$ > 7.25 $\times$ 10$^{14}$ $M_{\odot}$).
\end{itemize}
Four clusters are in common between these two sub-samples.

The \xmm\ observations are characterised by an exposure time which ensures a $S/N$=150 within $R_{500}$ in the [0.3-2.0] keV band. This condition has been requested to estimate the temperature profile at least up to $R_{500}$ (with a precision of $\pm$ 15\% in the region [0.8-1.2] $R_{500}$) and to obtain a measurement of both the mass derived from the $Y_X$ mass proxy (\cite{Kratzov2006}, where $Y_{\rm X}$=$M_{\rm g,500}T_{\rm X}$, $M_{\rm g,500}$ is the mass of gas within $R_{500}$ and $T_{\rm X}$ is the spectroscopic temperature estimated in the range [0.15–0.75] $R_{500}$) with $\pm 2$\% of uncertainty and the mass derived from HE at $R_{500}$ with $\sim$ 15 - 20\% precision level. For more details on the sample, on the scientific goals and on the strategy to observe homogeneously these systems in X-ray and to follow them up in other wave-bands, we refer the reader to \cite{CHEXMATE}.

\subsection{Preparation of the X-ray images}
Images were produced using the pipeline developed during the XMM-Newton Cluster Outskirts Project \citep[X-COP,][]{xcop,Ghirardini2019} and adopted by the CHEX-MATE collaboration. In particular, the {\it XMM-Newton} data were processed using the SAS software (version 16.1.0) and the extended source analysis software (ESAS) package \citep{Snowden2008}. Count-images, exposure maps, and particle background maps are extracted in the narrow [0.7-1.2] keV band, where the ratio between the source and background emission is maximised and, consequently, the systematics related to the subtraction of the EPIC background are minimised \citep{Ettori2010}. A detailed description of the procedure adopted will be presented in Bartalucci et al. 2022 (in prep.) 
and a complete gallery of the images is shown in Fig.~6 in \cite{CHEXMATE}. According to the CHEX-MATE pipeline, point sources in our observations are identified using the SAS tool \texttt{ewavelet} in two bands ($0.5-2$ keV and $2-7$ keV). Furthermore, we applied a filter in the $LogN-LogS$ distribution as described in \cite{Ghirardini2019}, to ensure a uniform level of the Cosmic X-ray Background (CXB) emission across the field of view. For the estimation of morphological parameters, we further inspected images by eye in order to identify residual point sources, which could affect our measurement. We masked identified point sources and filled the holes using an interpolation with the surrounding pixels. We point out that only point sources were masked, and thus substructures related to major or minor mergers are still present in the images, allowing a correct identification of the clusters dynamical state. The pixels size is 2.5 arcsec.   


\section{Initial visual classification}\label{visual classification}
\begin{figure}
    \hspace{-0.3cm}
    \includegraphics[scale=0.65]{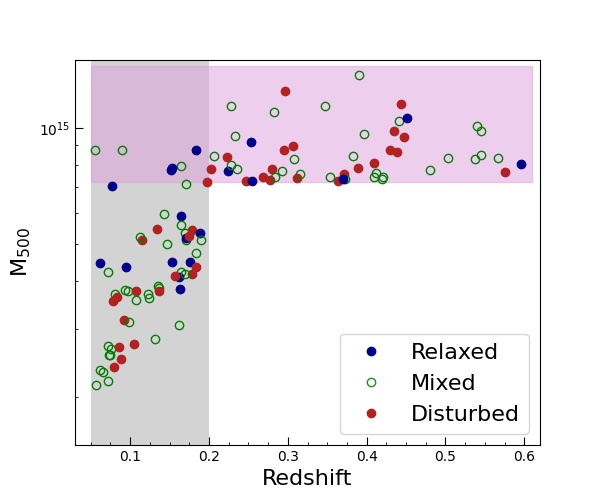}
    \caption{Distribution in the $M_{500}-z$ plane of the 118 CHEX-MATE clusters. The color scale represents the dynamical state obtained from the visual classification. The shaded area indicate the Tier 1 and Tier 2 redshift ranges in grey and violet, respectively. }
    \label{fig:classification_final}
\end{figure}

To study the link between the morphological parameters and the dynamical state of clusters, we first realised a visual classification of the objects of the CHEX-MATE sample. In particular, a group of seven X-ray astronomers inspected the images,  and rated the relaxation state of the clusters with a grade that ranges from 0 (most relaxed, i.e.  circular X-ray isophotes and without substructures) to 2 (most disturbed, i.e. double or complex objects with clear sign of merging). Results were then averaged and objects with rounded values equal to 0 were classified as relaxed (R, 19 clusters), objects with rounded values equal to 2 as disturbed (D, 37 clusters), objects with rounded values equal to 1 as Mixed (M, 62). An example of relaxed, mixed and disturbed system, as identified by the visual classification, is shown in Fig. \ref{fig:example_rmd}, while in Fig. \ref{fig:classification_final} we report the final dynamical state as a function of the redshift and the mass. It is possible to notice that the majority of the relaxed objects is located at redshift lower than $z$ < 0.25 and has a mass higher than $\sim 4 \cdot 10^{14} M_{\odot}$. No particular trend is instead observed for the disturbed and mixed class. The dynamical state obtained from this first analysis and the related uncertainty are reported in Table \ref{Tab:final_results}, column 'visual' (Appendix \ref{Appendix:parameters_values}).

\section{Morphological parameters}\label{parametri}

In this Section, we introduce the methods for the substructure and morphology characterisation of the CHEX-MATE objects. To estimate the morphological parameters listed below we considered a circular region within R$_{500}$ centred on the cluster X-ray peak.
This choice was adopted to avoid the contamination of signatures related to accretion processes, which are expected at larger radii \citep[e.g.,][]{Roncarelli2006}. Hydrodynamical simulations \citep[e.g.,][]{DeLuca2021} and observations \citep{Ghirardini2019,Eckert2014} indeed, show that within $R_{500}$ clusters are relatively relaxed unless a merger event modifies the existing conditions. Furthermore, the CHEX-MATE observation strategy, provides a coverage of this area for all the clusters of the sample \citep{CHEXMATE}. For these reasons we considered $R_{500}$ the optimal assumption to obtain a complete view of the dynamical state of clusters. The clusters centres are set at the brightest pixel of a Gaussian-smoothed ($\sigma \sim$ 15 pixels) background-subtracted and exposure corrected surface brightness image. The value of $R_{500}$, was derived from the \planck\ PSZ2 masses \citep{Planck2016}. 
In our analysis seven morphological parameters were taken into consideration: the light concentration, $c$, the centroid shift, $w$, two power ratios, the asymmetry, $A$, the smoothness, $S$, and the ellipticity, $\eta$. It results that $A$ and $S$ are not robust parameters, since they are strongly influenced by the signal-to-noise ratio (see Appendix \ref{Other_parameters} for the discussion). For this reason, we decided to threat their analysis in the Appendix, excluding them from the main text. Also the analysis of the ellipticity is discussed in Appendix \ref{Other_parameters}, since its behaviour reproduces the one of the quadrupole power ratio $P_2$ (a measure of the ellipticity of clusters, see the following subsections for more details).
\begin{figure*}
    \hspace{-0.3cm}
\includegraphics[scale=0.49]{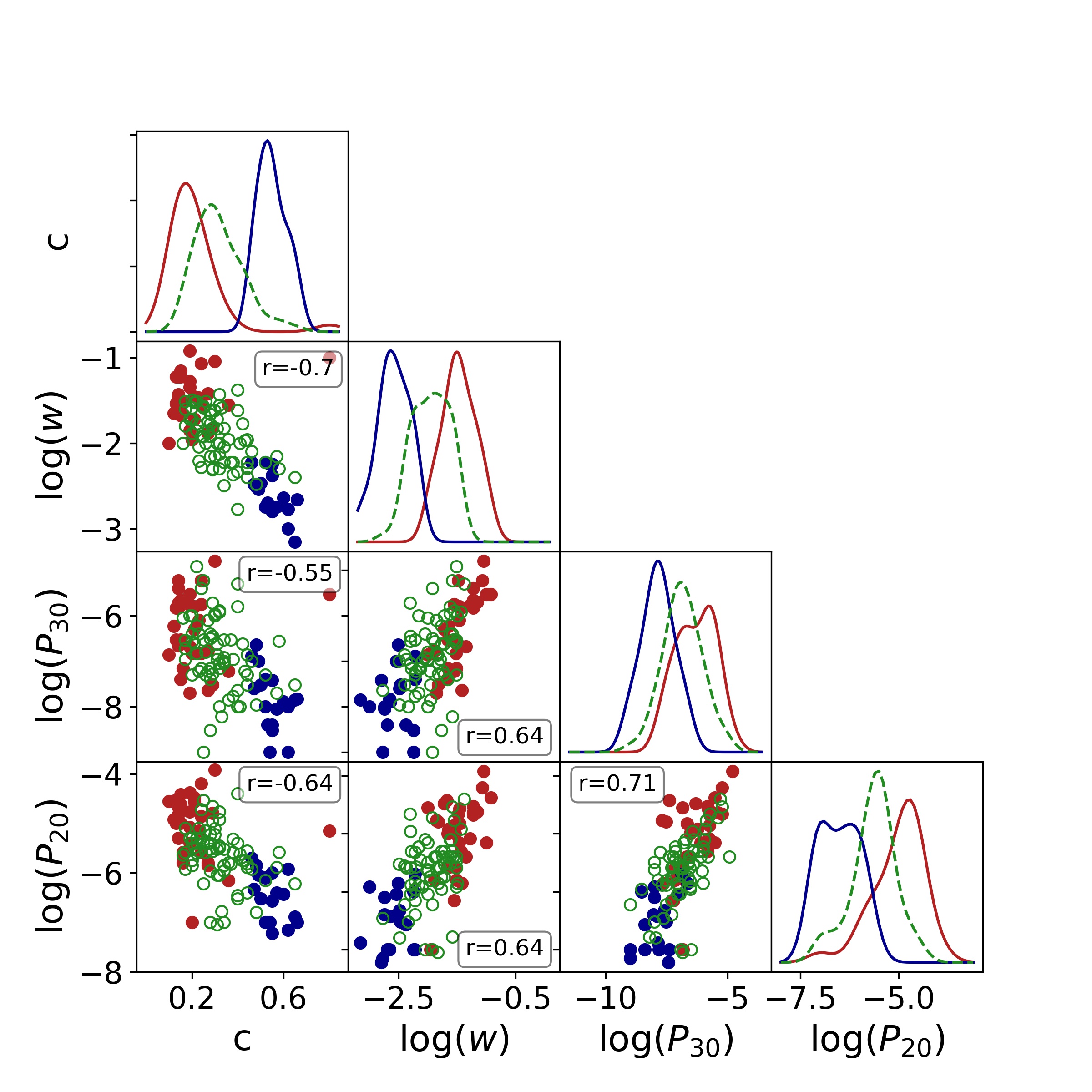}
\hspace{-0.7cm}
\includegraphics[scale=0.49]{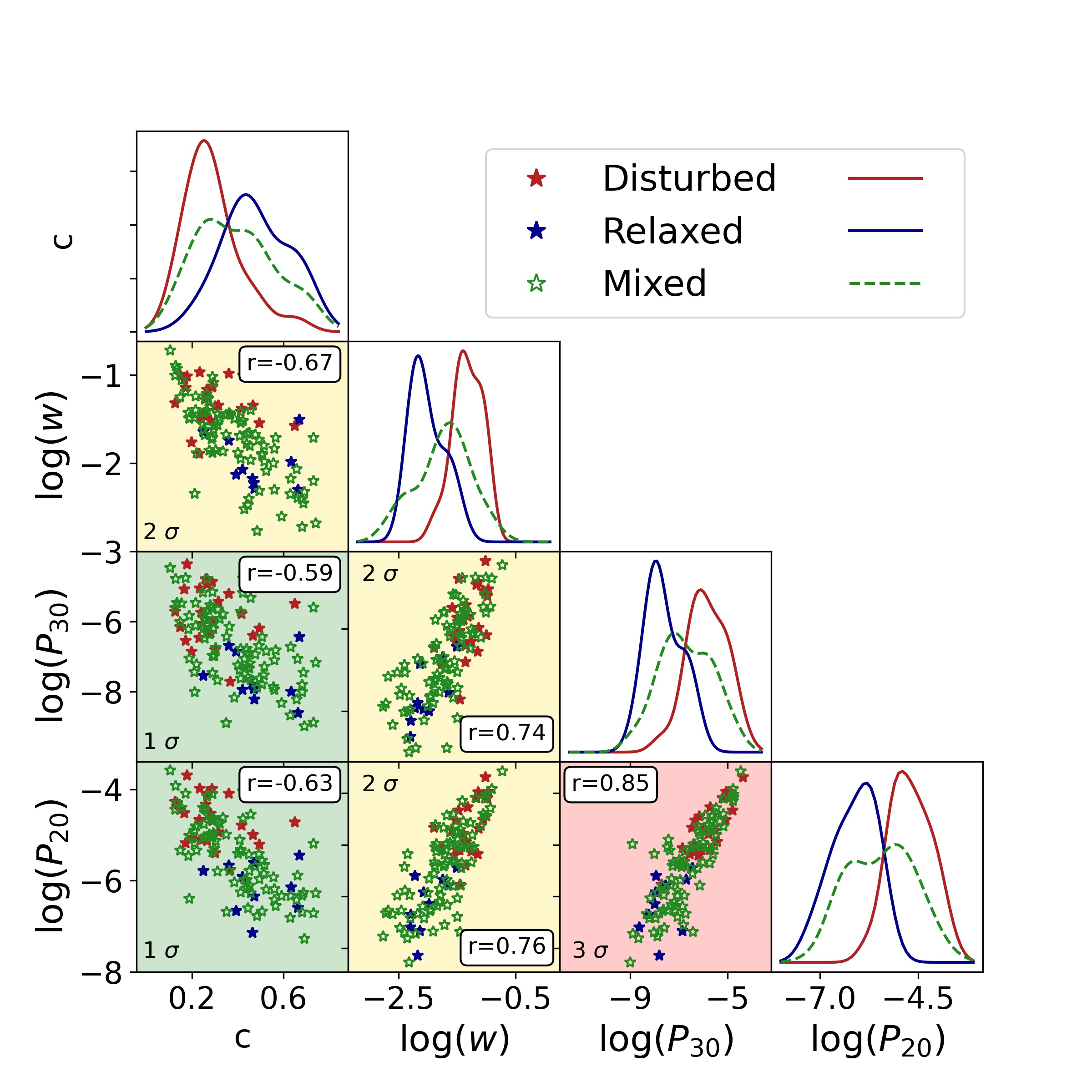}
    \caption{Left: distributions of the CHEX-MATE morphological parameters.  The different colours represent the dynamical state obtained from the visual classification (see Sect. \ref{visual classification}). Right: distributions of the simulated morphological parameters. The different colours represent the dynamical state obtained using the dynamical indicator $\chi$ (see Sect. \ref{simulated_sample}). The background colours of the plots represent the level of agreement with the observations: green is for 1 $\sigma$, yellow stays for 1-3 $\sigma$ and red for a level of agreement over 3 $\sigma$. The $r$ values reported in the boxes are the Spearman coefficients of the considered parameter pairs.} 
    \label{morphology_oss}
\end{figure*}
\subsection{Light concentration}\label{concentration}

The light concentration parameter (hereinafter, concentration), $c$, is defined as the ratio of the surface brightness (SB) inside two concentric apertures. It was introduced by \cite{Santos2008} to identify cool-core clusters at high redshift, using two apertures at 40 and 400 kpc, chosen to maximise the separation of concentration values between cool-core (CC) and non cool-core (NCC) clusters in the sample analysed there. However, the cluster volume enclosed in a fixed aperture evolves significantly with redshift and this behaviour can affect the selection of relaxed and disturbed systems \citep{Hallman2011}. Since our sample spans a large redshift range, we chose to define the concentration parameter as a function of the overdensity radii using the following apertures:
\begin{equation}
     c=\frac{\text{NC }(r< 0.15 R_{500})}{\text{NC }(r<R_{500})}.
\end{equation}
where $NC$ is the number of counts corrected for the exposure map measured in the considered aperture. On the basis of the above definition, the concentration computed from images is not corrected for the point spread function (PSF). The effects of this choice are that more distant objects are characterised by systematically lower concentrations than the low-redshift clusters, because more photons located in the centre are spread out across larger regions. The implication of our choice will be better discussed in Sect. \ref{concentration_discussion}.

\subsection{Centroid shift}

The centroid shift parameter \citep{Poole2006,Maughan2008}, $w$, is defined as the standard deviation of the projected separation between the X-ray peak and the centroid of the X-ray surface brightness computed within N (=10 in our case) apertures of increasing radius:
\begin{equation}
    w= \frac{1}{ R_{500}} \biggl[ \frac{1}{N-1}\sum_i (\Delta_i-\Bar{\Delta})^2 \biggr]^{\frac{1}{2}}
\end{equation}
where $\Delta_i$ is the distance between the X-ray peak and the centroid of the i-th aperture, and $R_{500}$ is the radius of the largest aperture. This parameter is useful for the characterisation of
the dynamical state of clusters because it is sensitive to the presence of X-ray bright clumps and substructures, which can produce significant changes on the X-ray centroid.

\subsection{Power Ratios}

The power ratio parameters were first introduced by \cite{buote1995} and are based on the idea that the X-ray surface brightness of a cluster could be the representation of its projected mass distribution. They are computed as a multipole decomposition of the two-dimensional projected mass distribution inside a certain aperture, but instead of the mass, the X-ray surface brightness is used. The $m$-order ($m>$0) power ratio is defined as $P_m/P_{0}$, with: \begin{equation}
P_0=[a_0 \text{ln}(R_{500})]^2,    
\end{equation}
and
\begin{equation}
    P_m=\frac{1}{2m^2 R_{500}^{2m}}(a_m^2+b_m^2)
\end{equation}
where $a_0$ is the total intensity within the aperture radius $R_{500}$ and the moments $a_m$ and $b_m$ are calculated by:
\begin{equation}
    a_m(R)= \int_{R<R_{500}} S(x) R^m \cos(m \phi)d^2x
\end{equation}
and
\begin{equation}
    b_m(R)= \int_{R<R_{500}} S(x) R^m \sin(m \phi)d^2x
\end{equation}
where $S(x)$ is the surface brightness at the position $x$=($R$, $\phi$).
The quadrupole power $P_2$ quantifies the ellipticity of the clusters, $P_3$ informs about bimodal distribution and is the most useful to identify asymmetries or presence of substructures, $P_4$ is similar to $P_3$ but more sensitive to smaller scales (for this reason they are strongly correlated). In the following analysis we are going to focus on the ratios $P_2$/$P_0$ and $P_3$/$P_0$ (hereafter $P_{20}$ and $P_{30}$, respectively). This choice is due to the properties of $P_{30}$, which is one of the most unambiguous indicators of an asymmetric
cluster structure \citep{Jeltema2005,Jeltema2008,cassano2010,Weissmann2013,Rasia2013,Lovisari2017,Cialone2018}, and to our wish to investigate the power of a indicator of the ellipticity of clusters, such as $P_{20}$.\\ 

\section{Morphological analysis of the CHEX-MATE sample} \label{correlazioni}
Starting from the dataset and the methods introduced above, we carried out a morphological analysis of the CHEX-MATE sample. For each cluster,  we estimated the values of the four parameters presented in Sect. \ref{parametri} and their associated statistical errors through a Monte-Carlo 100 re-sampling of the counts per pixel of the original image according to their Poissonian error \citep[the technique is implemented in an IDL routine already used for the analyses presented in][]{cassano2010,Donahue2016,Lovisari2017}. 

We then investigated the presence of correlations between pairs of these parameters, by determining the Spearman rank correlation coefficient, $r$, a value that varies in the range -1 $\leq r \leq$ 1 assuming extreme values (-1 or 1) when each of the variables tested is a perfect monotone function of the other. The results of the analysis are reported in Fig. \ref{morphology_oss}, left panel. All the parameter pairs are characterised by significant correlations (i.e., $|r|>$0.5), with $P_{20} - P_{30}$ and $c$ -- $w$ showing the highest Spearman coefficient ($r$ = 0.71 and $r$ = -0.7, respectively). The corner-plot also shows the parameter distributions of the relaxed and disturbed systems, as defined by the visual classification (see Sect. \ref{visual classification}). We found that, clusters with different dynamical states are located in distinct regions of the parameter-parameter planes; for example, in the $c$ -- $w$ plot, relaxed clusters occupy the lower right region, while disturbed systems are placed in the upper left region. It is thus possible to identify thresholds above (or below) which clusters can be classified as relaxed (or disturbed) and vice versa, depending on the behaviour of the considered morphological parameter. As highlighted by the density plot of Fig. \ref{morphology_oss} (left panel), the concentration and the centroid shift are confirmed to be powerful identifier of the relaxed and disturbed populations \citep{Santos2008,cassano2010,Hudson2010,Lovisari2017}: the distribution of the values of these two classes of objects are indeed well separated. Also the power ratios are able to distinguish relaxed and disturbed systems. However, it is possible to observe an overlap of the distributions of the values of these parameters, thus suggesting that by defining a threshold for the identification of the dynamical state, some clusters may be erroneously classified as relaxed instead of disturbed and vice versa. Finally, for all the four parameters, the systems classified as "mixed" represent an intermediate class between the relaxed and disturbed populations. 

\section{Systematics in the measurements of morphological parameters}

\begin{figure*}
\hspace{-0.7cm}
\begin{subfigure}{0.475\textwidth}
\includegraphics[scale=0.7]{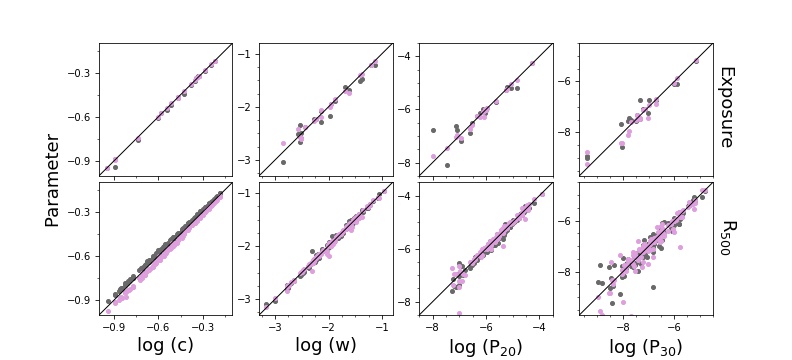}
\end{subfigure}
\caption{First row: Comparison between the parameters estimated from the original images (x-axis) and the parameters estimated from images with halved $t_{exp}$ (violet) and with $t_{exp}$ = 5 ks (grey). Second raw: comparison between the parameters estimated using $R_{500}$ (x-axis) and the parameters computed in a region of $r =$ 0.95 $R_{500}$ (violet) and $r =$ 1.05 $R_{500}$ (grey).}
\label{fig:systematics_50p}
\end{figure*}

\begin{table*}
\centering
\renewcommand{\arraystretch}{1.3}
\caption{Medians and interquartile ranges of $\Delta$.}\label{dispersion_median}
\begin{tabular}{|l|cc|cc|cc|cc|}
\hline
 & \multicolumn{2}{c|}{c}  & \multicolumn{2}{c|}{w} & \multicolumn{2}{c|}{$P_{20}$}  & \multicolumn{2}{c|}{$P_{30}$} \\  \cline{2-9}
 &    $\Delta_c$        &   IQR$_c$       &       $\Delta_w$      &   IQR$_w$       &  $\Delta_{P_{20}}$         &        IQR$_{P_{20}}$  &   $\Delta_{P_{30}}$        & IQR$_{P_{30}}$         \\ \hline
$t_{{\mathrm{exp, 5 ks}}}$ &  0.004         &   0.007      &    0.06       &   0.08       & 0.10          & 0.3         &      0.20     &     0.3     \\
$t_{{\mathrm{exp, 50 \%}}}$   &  0.003         &  0.003        & 0.03          &   0.07       &  0.05         &    0.09      &   0.12        &  0.2        \\
$R_{500 + 5\%} $  & 0.02          &  0.01        &    0.02       &  0.01        & 0.06          &   0.07       &    0.14       &   0.20       \\
$R_{500 - 5\%}$  &      0.02     &  0.01        & 0.02          &    0.02      &      0.07     &    0.07      &         0.12  &       0.19  \\ \hline 
\end{tabular}
\tablefoot{In each row are reported the values of $\Delta$ and of the interquartile ranges obtained from the tests presented in Sect. \ref{exposure} and \ref{radius} for the four morphological parameters. }
\end{table*}

In this Section, we test the robustness of the morphological parameters, by investigating whether their behaviour is influenced by the quality of the images used or by the assumptions/criteria adopted in our analysis. In particular, we investigated how the following systematic effects can alter our analysis:
\begin{enumerate}
    \item exposure time;
    \item region considered for the estimation of the morphological parameters.
\end{enumerate}
\subsection{Effects of the exposure time}\label{exposure}
The CHEX-MATE observations are characterised by high quality. To understand the robustness of our analysis, we tested the stability of the morphological parameters when images with lower exposure times are used. To this aim, we considered a sub-sample of 20 CHEX-MATE clusters (5 Relaxed, 5 Disturbed and 10 Mixed as defined by the visual classification) and we repeated the morphological analysis of the X-ray images now selected with two different exposure times, one which halved the total exposure time, $t_{{\mathrm{exp}}}= 0.5 \times t_{\mathrm{exp,total}}$, and the other with a minimal value of 5 ks.
We then computed the dispersion $\Delta$ following the relation:
\begin{equation}
   \Delta = |\log_{10}(\mathcal{P}_{red})-\log_{10}(\mathcal{P}_{or})|
\end{equation}
where $\mathcal{P}_{red}$ are the parameters computed using images with reduced exposure times and $\mathcal{P}_{or}$ are the parameters computed using the original images. The median values and the interquartile ranges (IQR) of $\Delta$ obtained from the two types of images are reported in Table \ref{dispersion_median} (first and second row). In Fig. \ref{fig:systematics_50p} (first row), we also plotted the values of the morphological parameters computed from the new images (violet, 50\% $t_{\mathrm{exp,total}}$, and grey $t_{{\mathrm{exp}}}$ = 5 ks) as a function of the values computed using the original ones. It is possible to observe a low scatter between the two types of estimations. 

\subsection{Effects of the assumed R$_{500}$}\label{radius}
As presented in Sect. \ref{parametri}, all the parameters are estimated inside a region of radius $r$ = $R_{500}$ derived from the \planck\ PSZ2 masses. However, uncertainties that may affect the measure of the \planck\ masses could influence also the estimation of $R_{500}$. The understanding of how this could impact our analysis is crucial. To investigate this point, we considered the CHEX-MATE clusters and we computed the morphological parameters  inside two circular regions of radius  $r \sim$ 1.05 $\cdot$ $R_{500}$ and 0.95 $\cdot$ $R_{500}$. Also in this case, we computed the dispersion $\Delta$:
\begin{equation}
   \Delta = |\log_{10}(\mathcal{P}_{R_{500}\pm 5\%})-\log_{10}(\mathcal{P}_{R_{500}})|
\end{equation}
where $\mathcal{P}_{R_{500}\pm 5\%}$ are the parameters computed using images with increased or decreased radius and $\mathcal{P}_{R_{500}}$ are the parameters computed using the original aperture. In Fig. \ref{fig:systematics_50p} (second row), we show the scatter plot of the new parameters and the old ones, while in Table \ref{dispersion_median}(third and fourth row) we report the medians and the interquartile ranges of $\Delta$. We found that no significant difference is present when taking into account possible uncertainties related to the radius estimations. 

In addition to this test, we also compared our estimations of the morphological parameters with the ones obtained using a region of radius equal to 0.5 $\cdot R_{500}$. The change of the radius in which to compute the morphological parameters could indeed affect the classification of the dynamical state. The results of this analysis are reported in Appendix \ref{05analysis}.

\section{Construction of the $M$ parameter}\label{sect:Mparameter}

\begin{figure*}
    \centering
    \includegraphics[scale=0.6]{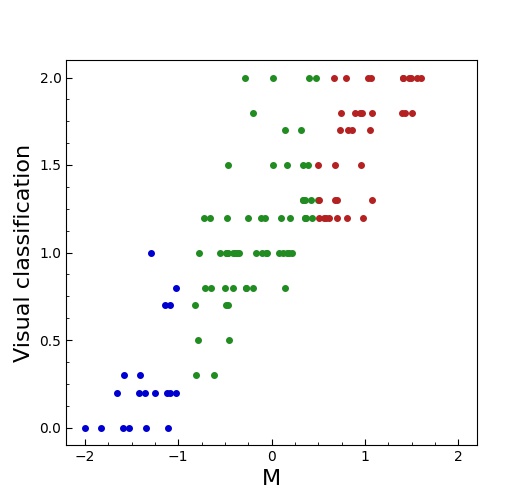}
    \includegraphics[scale=0.6]{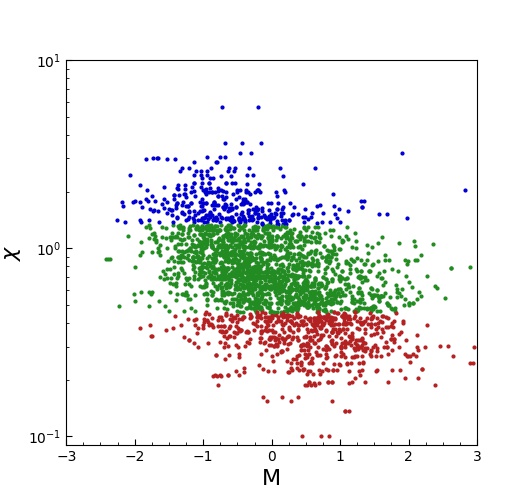}
    \caption{Left: Comparison between the CHEX-MATE dynamical state obtained from the visual classification and from the $M$ parameter. The colours represent the dynamical classification obtained on the basis of the $M$ parameter (blue for relaxed, red for disturbed and green for mixed clusters). Right: distribution of the simulated sample in the $\chi - M$ space. The colours represent the relaxed (blue), disturbed (red) and mixed (green) class defined on the basis of $\chi$ (see Sect. \ref{simulated_sample}).}
    \label{fig:dyn_M}
\end{figure*}

The test presented in the previous Section highlighted that the centroid shift, the power ratios and the concentration could be considered powerful and robust morphological indicators. Therefore, we decided to combine the information included in these parameters, in order to build a unique indicator of the grade of relaxation of a cluster. This new indicator allows us to establish a ranking of the dynamical state of the clusters of a sample, which can be used to to identify the population of the most relaxed and most disturbed objects. The definition of this new quantity is \citep[see e.g.][]{Rasia2013,Cialone2018}:
\begin{equation}
    M=\sum \frac{\log_{10}(\mathcal{P}^{\alpha_{\mathcal{P}}})-<\log_{10}(\mathcal{P}^{\alpha_{\mathcal{P}}})>}{\sigma_{log_{10}(\mathcal{P}^{\alpha_{\mathcal{P}}})}}
\end{equation}
where $\sigma$ is the standard deviation of the considered parameter, $\mathcal{P}$, and the term $\alpha_{\mathcal{P}}$ is considered equal to -1 only in the case of the concentration, otherwise it is fixed equal to +1. For each cluster, $M$ represents the sum of the differences of the four parameters for the mean of their distributions, normalised by their standard deviations. The log-scale was introduced to take into account the different ranges of values covered by the four morphological parameters.  According to this definition, relaxed systems are expected to show low values of $M$, while disturbed systems should be characterised by high values. Furthermore, it appears clear, that the $M$ parameter does not provide an "absolute" grade of relaxation, but just a relative value based on the analysed sample: i.e. it ranks the clusters from the most relaxed to the most disturbed, given the distribution of the parameters of the sample. 

We estimated $M$ for the objects of the CHEX-MATE sample and we identified the most relaxed and disturbed systems of the sample using as reference the visual classification presented in Sect. \ref{visual classification}. In particular, using the fractions of relaxed and disturbed systems obtained by that classification, we verified whether the first 19 (or 37) objects with the lowest (or highest) values of M are effectively classified as relaxed (or as disturbed) by the visual classification. If that is the case, we refer to correct detection, $C$, while if the objects are classified as relaxed (or disturbed) by $M$ and as disturbed (or relaxed) by the visual classification we refer to wrong detection, $W$. We found a high efficiency of $M$ in identifying the most relaxed and disturbed systems, with correct detection equal to $C_R$ = 79 \% and $C_D$ = 70\% and no wrong detection ($W_R$ and $W_D$ equal to zero). This means that the not-correct detection are mostly due to the mixed population. A comparison between the $M$ and visual classifications is shown in Fig. \ref{fig:dyn_M}, left panel. In Appendix \ref{PCA_appendix}, we report the results obtained from the Principal Component Analysis (PCA), another method to combine the information included in the four morphological parameters.

\section{Discussion}

\subsection{Distribution of the morphological parameters}\label{Sect:distributions}

\begin{figure*}
\hspace{-1.5cm}
   \includegraphics[scale=0.55]{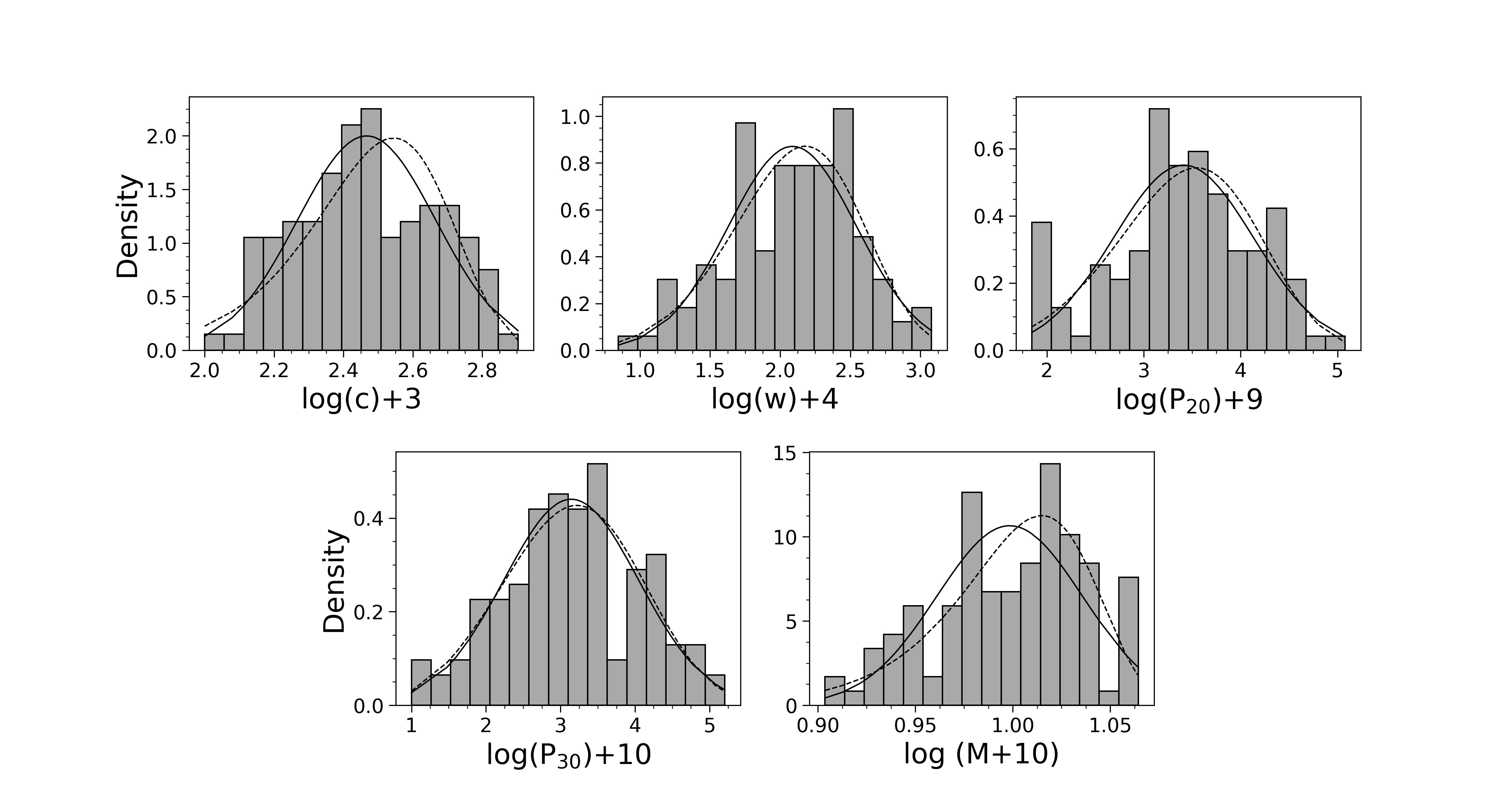}
   
    \caption{Distribution of $c$, $w$, $P_{20}$, $P_{30}$ and $M$. The solid line represents the model obtained from the clustering analysis, while the dashed one represents the Weibull function. The parameters $c$, $w$, $P_{20}$ and $P_{30}$ were multiplied by 10$^3$, 10$^4$, 10$^9$ and 10$^{10}$, respectively. For the $M$ parameter instead, we added 10 to the original values. }
    \label{fig:total_G}
\end{figure*}

In this Subsection we investigate the presence or absence of bimodality in our parameter distributions. In particular, we performed a maximum likelihood fit of our unbinned data using the normal mixture model, and a commonly used positively skewed function: the Weibull distribution.

For this part of the analysis, we resort to  the \texttt{MCLUST} \citep{Fraley2002,Mclust} package and the \texttt{FITDISTR} and \texttt{FITDIST} function of the \texttt{MASS} \citep{venables2013} and \texttt{fitdistrplus} \citep{FITDISTRPLUS} packages in the software environment R, version 3.6.3 \citep{R}. \texttt{MCLUST} allows to perform a cluster analysis (or 'clustering'), which consists in grouping together similar set (or cluster) of objects of a dataset. The objects of each cluster are comparatively more similar to objects of that group than those of the other clusters. Clustering is a main task of exploratory data analysis (EDA) and provide us the opportunity to see what the data can tell us beyond the formal modelling or the a priori hypothesis. The \texttt{MCLUST} method realises a maximum likelihood fit assuming that a number from 1 to 9 normal components are present in the data. Also the function \texttt{FITDISTR} performs a maximum likelihood fit of the data using some probability distribution functions, either calculated using analytic formulae (as e.g. in the log-normal case) or computed by optimisation of the likelihood. This method was used to fit our data with the Weibull distribution. 

To compare and select the most appropriate models for the description of the empirical dataset, we considered the Bayesian information criterion \citep[BIC,][]{Schwarz1978}
defined as BIC $=$ 2$\ln (L) - k \log(n)$, where $L$ is the likelihood, $k$ is
the number of parameters of the model and $n$ is the number of data
points; $k\log(n)$ is the penalty term that compensates the difference
in likelihood due to an increase in the number of fitting parameters. The best model is the one that maximizes the BIC. For the interpretation of the differences between the BIC values obtained with different models we considered the following commonly adopted thresholds: a BIC difference of 0–2 is a weak confirm, 2–6 a positive confirm, 6–10 a strong confirm and $>$10 a very strong confirm of the model with the greater BIC value \citep{Kass1995,Raftery1995}. 

Since the Weibull function is not defined for negative values, we applied a normalisation to the morphological parameters. In particular, we multiplied by 10$^3$, 10$^4$, 10$^9$ and 10$^{10}$, $c$, $w$, $P_{20}$ and  $P_{30}$ respectively, and we added 10 to the $M$ values. In this way we could consider their log distributions. The results of our analysis are shown in Fig.~\ref{fig:total_G} and Table~\ref{BIC_values}. 
For all the five parameters, the clustering analysis revealed the presence of a single component. For this reason, the BIC values reported in Table \ref{BIC_values} are related to a single Gaussian component model. We concluded that there are no parameters showing signs of bimodality and therefore, that the cluster population cannot be easily divided in two populations. A similar result was already observed for $c$. $P_{20}$ and $P_{30}$ in a recent analysis on the eFEDS sample \citep{Ghirardini2021}. Given the BIC values obtained, it is not possible to unambiguously identify the model that best fits our data. In particular, for $w$, $P_{20}$ and $P_{30}$ the discrepancy between the BIC values is $\sim$ 2, and represents only a weak confirm for the model with the highest BIC value (i.e. Weibull for $w$, $P_{30}$ and $M$, and Gaussian for $P_{20}$). For what concerns the concentration, the discrepancy between the BIC values is 5 and could be considered a positive confirm for the single Gaussian component model. We thus concluded that the distribution of the concentration is log-normal. A similar behaviour was already found both in the \planck\ selected sample analysed in \cite{Rossetti2017}, where a log-normal best fit for $c$ contrasted the bimodal best-fit distribution of the X-ray selected sample ME-MACS, and  the eFEDS sample analysed in \cite{Ghirardini2021}. In this latter analysis, also the distributions of $P_{20}$ and $P_{30}$ were taken into consideration. As in our case, it was not possible to unambiguously identify the best fit model for these parameters, but skewed and log-normal distributions were preferred.

An analogous behaviour was observed, by repeating this analysis on the Tier 1 and Tier 2, separately. The results are shown in Table \ref{BIC_values}. Also in this case the BIC values reported for the Gaussian mixture models are related to a single Gaussian component function, which results to be the best one in reproducing the distributions of the five parameters. Also in this case we found no sign of bimodality neither in the uniform low-$z$ sub-sample nor in the uniform high mass sub-sample. However, since the BIC values are very similar, it is not possible to derive which is the best fit model for our data.  

\begin{figure*}
    \hspace{-1.5cm}
    \includegraphics[scale=0.7]{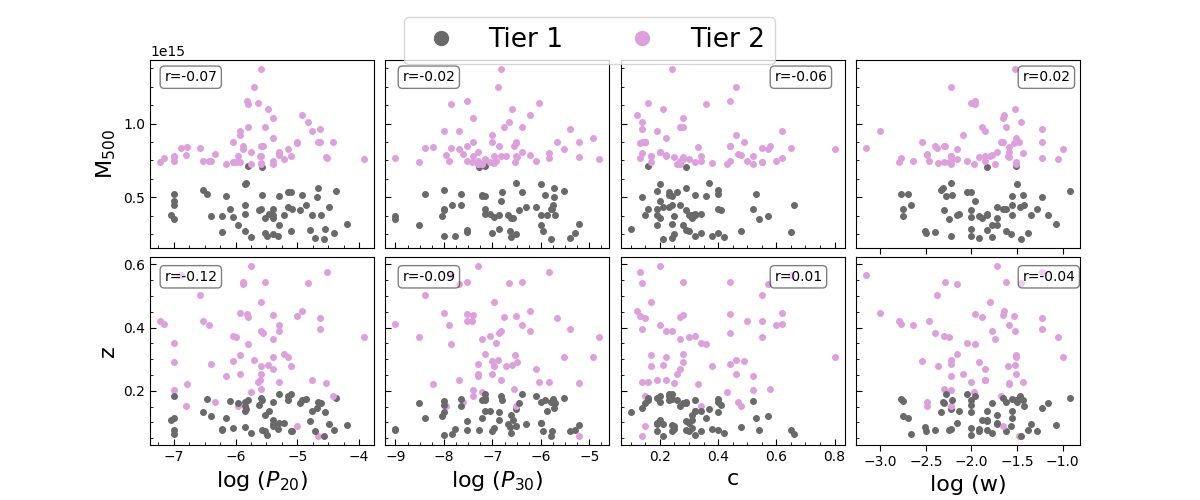}
    \caption{Mass and redshift distributions of the morphological parameters. The different colours represent the Tier 1 (grey) and Tier 2 (violet) objects. In the boxes is reported the Spearman coefficient, $r$.}
    \label{fig:Mz_corr}
\end{figure*}

\begin{table*}
\caption{BIC values obtained from the fit realised on our data, using the Gaussian mixture model or the Weibull function.} \label{BIC_values}
\centering
\renewcommand{\arraystretch}{1.2}
\begin{tabular}{|c|ccccc|ccccc|ccccc|}
\hline 
                 & \multicolumn{5}{c|}{CHEX-MATE} & \multicolumn{5}{c|}{Tier 1} & \multicolumn{5}{c|}{Tier 2} \\ \cline{2-16} 
                 & $c$   & $w$   & $P_{20}$   & $P_{30}$  & $M$  & $c$  & $w$  &$P_{20}$  & $P_{30}$  & $M$  & $c$  & $w$  & $P_{20}$  & $P_{30}$  & $M$  \\ \hline 
\begin{tabular}[c]{@{}c@{}}Gaussian \\ (1 component) \end{tabular} &  36   & -160     & -267      &  -321    &    431 & 6   &    -81 & -127     & -159     & 208    & 23   &-78    & -135     & -153     &  205  \\ \hline
Weibull          & 31.0     & -157.3    & -269.2      & -320.3      & 433.4   & 4   & -81   & -127     &    -163  & 208   & 19    & -75   & -132     & -153     &  209 \\ \hline
\end{tabular}
\tablefoot{The models were applied to the logarithmic distributions of the morphological parameters. Since the Weibull function is not defined for negative values, $c$, $w$, $P_{20}$ and $P_{30}$ were multiplied for 10$^3$, 10$^4$, 10$^9$ and 10$^{10}$, respectively. For the $M$ parameter instead, we added 10 to the original values (see Sect. \ref{Sect:distributions}). Besides the results arising from the analysis of the entire CHEX-MATE sample, we also present the results obtained for the Tier 1 and Tier 2, separately.}
\end{table*}
Moreover, we investigated the distributions of the morphological parameters with respect to two intrinsic properties of clusters: the mass and the redshift. As is it possible to observe from Fig. \ref{fig:Mz_corr}, no specific trend is observed and the low values of the Spearman coefficients highlight the absence of correlations. This result is in agreement with previous findings by \cite{Boehringer2010, mantz2015,Lovisari2017,Rossetti2017}.

\subsection{Final CHEX-MATE classification}\label{final_classification}
In Fig. \ref{fig:dyn_M} (left panel), we presented a comparison between the dynamical state derived from the visual classification (y-axis) and the one derived on the basis of the $M$ values (colour-scale). As already highlighted by the quantities $C_R$ and $C_D$ in Subsect. \ref{sect:Mparameter}, some clusters may be classified as mixed instead of relaxed/disturbed by one of the two classification. In order to investigate the differences between the two classifications, we inspected by eye the objects for which an agreement was not obtained. For example, we verified that some objects classified as relaxed by $M$, show not only a centrally peaked emission but also some subtructures in their outskirts. Due to this feature these systems were classified as mixed by the visual inspection. In order to obtain a classification of the sample as accurate as possible, we decide to define as relaxed (or disturbed) only the clusters for which the two classifications provide the same results. Therefore, our analysis identified 15 relaxed and 25 disturbed clusters. To have an overview of the dynamical state of the entire sample, we realised a continuous classification using the following criteria. The first 15 and the last 25 objects of the continuous classification are respectively the most relaxed and disturbed systems identified by both the visual and $M$ classification. All the other systems are classified as mixed. Inside these three populations, clusters are ranked on the basis of $M$ (from the lowest to the highest values). The final rank is presented in Table \ref{Tab:final_results} (Appendix \ref{Appendix:parameters_values})

\subsection{Comparison with other observed samples}\label{comparison_oss}

\begin{figure*}
\centering
    \includegraphics[scale=0.5]{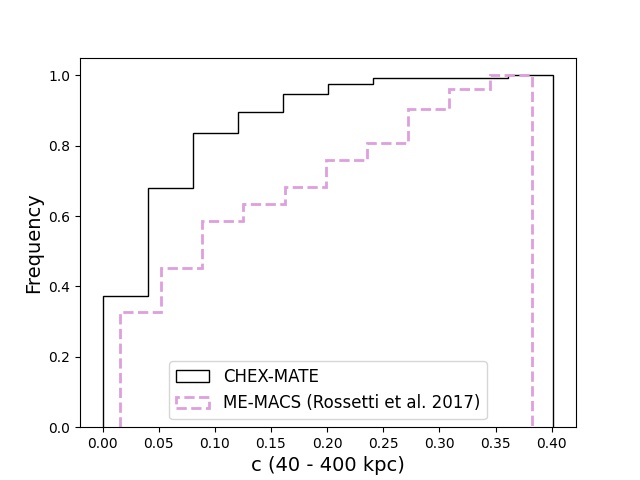}
    \includegraphics[scale=0.5]{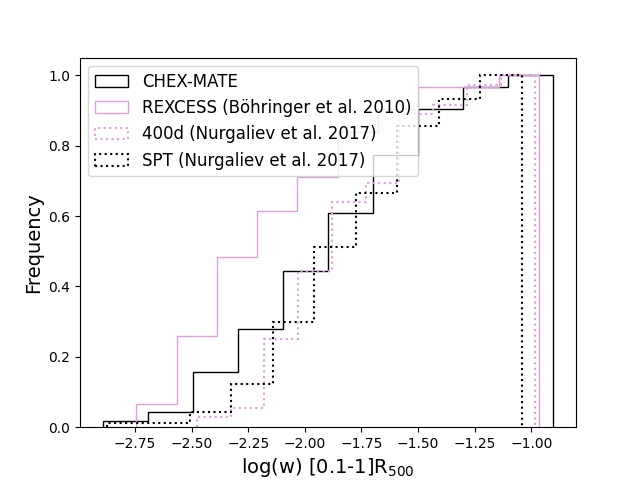}
    \caption{Left: comparison between the concentration values of the CHEX-MATE and ME-MACS sample. Right: comparison between the centroid shift values of the CHEX-MATE, REXCESS, 400d and SPT samples.}
    \label{fig:comparison}
\end{figure*}

The characterisation of the dynamical state of cluster samples has been extensively explored. However, it is not easy to compare the results obtained from different analysis. The definition of the morphological parameters is often different and it depends on the goals or the limitation of the analysis. In addition to this, we have also to consider that selection effects may affect the definition of cluster samples. In particular, many studies have highlighted the presence of discrepancies between SZ and X-ray samples: the latter show typically a higher fraction of cool core systems (i.e., systems with a centrally peaked emission usually defined as relaxed), while the SZ-selected clusters are characterised by a higher fraction of substructures than the X-ray selected systems \citep{Rossetti2017}. However, not all the analysis converge towards this result. For example, using the photon asymmetry and the centroid shift parameters, \cite{nurgaliev2017} found no significant statistical difference between the X-ray morphology of X-ray and SZ-selected samples, suggesting that the two are probing similar populations of clusters. A comparison between the X-ray morphology of the CHEX-MATE clusters and other samples analysed in literature may help us figure out, how the CHEX-MATE sample is able to represent the underlying cluster population.

To this aim, we compared CHEX-MATE with three X-ray selected samples and one SZ-selected sample: 
\begin{itemize}
    \item an extended version of the MACS sample described in \cite{Mann2012} (ME-MACS hereafter). As MACS, this sample is build from the RASS Bright
    Source Catalogue \cite{voges1999}, with a flux limit $f_{RASS}$[0.1-2.4] KeV $>$ 1 $\cdot$ 10$^{-12}$ erg cm$^{-2}s^{-1}$. In contrast to MACS, which is composed of the most distant systems ($z$ > 0.3), ME-MACS includes also lower redshift clusters ($z$ > 0.15)  and has an additional luminosity cut $L_{RASS}$[0.1–2.4 keV] > 5 $\cdot$ 10$^{44} erg \, s^{-1}$. These features make the ME-MACS sample a purely X-ray-selected sample, based on a flux-limited survey whose distribution is similar to the CHEX-MATE redshift distribution.
    \item the REXCESS sample, \citep[e.g., The REpresentative XMM-Newton Cluster
Structure Survey][]{Bohringer2007}, which is a  a representative and statistically unbiased subsample of 33 galaxy clusters extracted from the REFLEX cluster catalogue with a rigorous selection in the luminosity–redshift space \citep[see details in][]{Bohringer2007}.
\item  high-$z$ part of the ROSAT PSPC 400 deg$^2$ cluster survey \citep{Burenin2007}, abbreviated hereafter as 400d, for our X-ray-selected sample. This sample consists of 36 clusters in the redshift range of 0.35$<z<$ 0.9 and the mass range of 10$^{\mathbf{14}}$ $M_{\odot}<M_{500}<$ 5 $\cdot$ 10$^{14} M_{\odot}$. 

\item  the 2500 deg2 SPT survey of \cite{Bleem2015}. This sample is composed of 90 clusters which are among the most massive of the SPT-selected
clusters. The systems span the redshift range of 0.25 $<z<$ 1.2 and the mass range range 2$\cdot$10$^{\mathbf{14}}$ $M_{\odot}<M_{500}<$ 2 $\cdot$ 10$^{15} M_{\odot}$
\end{itemize}
The parameters used for this comparison are the concentration estimated by \cite{Rossetti2017} for the ME-MACS sample, the centroid shift obtained by \cite{Boehringer2010} for the REXCESS sample and the centroid shift obtained by \cite{nurgaliev2017} for the 400d and SPT samples. Using the procedure adopted in our analysis, we recomputed $c$ and $w$ following the definition used in those studies. In particular: for the concentration we adopted two apertures of radius equal to 40 and 400 kpc, for the centroid shift of \cite{Boehringer2010} we used a region included between [0.1-1] $R_{500}$. For the centroid shifts of \cite{nurgaliev2017}, we used the definition of \cite{Boehringer2010}, and we scaled the values by a factor of 1.5. This normalisation is required because the definition of $w$ presented in \cite{Boehringer2010} (which is the same adopted in this paper), differs from the one of \cite{nurgaliev2017}. According to this latter $w$ is the squared difference between the position of the centroid and the average position of the centroid. The value of the normalisation (1.5) was already determined in \cite{nurgaliev2017}. We found that CHEX-MATE has a higher fraction of objects with low concentration compared to ME-MACS and a lower fraction of objects with lower values of the centroid shift. These results are confirmed by the KS-test, which return in both cases a p-value lower than 0.01. For what concerns 400d, we found that, despite being a X-ray selected sample, it show a behaviour similar to the SZ-selected samples (CHEX-MATE and SPT). The differences observed between this sample and other X-ray-selected samples have been already debated in the literature \citep[e.g.,][]{Santos2010,Ma2011,mantz2015,Rossetti2017} and are related to the detection procedures adopted, which made 400d rather unique
among X-ray samples.

\subsection{Comparison with simulations}
In this Subsection we present the result of the comparison between the CHEX-MATE observations and the simulations. The simulated sample at our disposal is provided by {\sc The Three Hundred}\footnote{https://the300-project.org} collaboration \citep{Cui2018} and is composed of 1564 objects spanning a wide range of redshift ($0 < z < 0.59$) and masses ($M_{500}$ > 1.1 $\times$ 10$^{14}$ h$^{-1}$ M$_{\odot}$). For each object three images related to three different orientations are provided, thus allowing to build a sample of 4692 (1564 $\times$ 3) different maps.

\subsubsection{\textbf{Dynamical state of the simulations}}\label{simulated_sample}
The additional information provided by the analysis of smoothed-particle hydrodynamics (SPH) simulations resides in the knowledge of the physical properties of each particle. In order to determine the dynamical state of simulated clusters, it is thus possible to take advantage of this, referring to quantities computed in 3D that would be unreachable with the current observational techniques. In this analysis, we characterised the grade of relaxation of the simulated sample by means of the following indicators:
\begin{itemize}
    \item  the mass fraction of all sub-halo in the cluster, $f_s$, where the sub-halos are identified with the Amiga Halo Finder, AHF3 \citep{Knollmann2009}, whenever the structure has at least 20 particles. This parameter is defined as:
    \begin{equation}
        f_s = \frac{\Sigma_i M_i}{M_{500}}
    \end{equation}
    where M$_{500}$ is the mass of the cluster enclosed in $R$ < $R_{500}$ and  $M_i$ is the mass of the sub-halos in the same volume;
    \item the offset of the centre of mass, $\Delta_r$, defined as:
    \begin{equation}
        \Delta_r = \frac{|r_{cm}-r_c|}{R_{500}}
    \end{equation}
    where $r_{cm}$ is the centre-of-mass position of the cluster and $r_c$ is the theoretical centre of the cluster, identified as the position of the highest density peak.
    \item the virial ratio $\eta$, based on the virial theorem and defined as:
    \begin{equation}
        \eta = \frac{2T-E_s}{|W|}
    \end{equation}
    where $T$ is the total kinetic energy, $E_s$ is the surface pressure energy
from both collisionless and gas particles, and $W$ is the total potential
energy \citep[see][, for more details]{Klypin2016,Cui2017,John2019}.
\end{itemize}
To obtain a continuous classification of the dynamical state of clusters it is possible to combine this indicators using the following relation \citep{Haggar2020}:
\begin{equation}
    \chi = \sqrt{\frac{3}{\biggl(\frac{\Delta_r}{0.04}\biggl)^2+\biggl(\frac{f_S}{0.1}\biggl)^2+\biggl(\frac{|1-\eta|}{0.1}\biggl)^2}}
\end{equation}
For a relaxed cluster, $\Delta_r$ and $f_s$ are expected to be minimal, and $\eta \rightarrow$ 1 \citep{Cui2017}. Therefore, they are expected to show high values of $\chi$. Unfortunately, in literature, there is not a unique selection of the thresholds to use to segregate among relaxed and disturbed clusters \citep[see also][]{Cui2017}. The variety of choices made by different authors is partially explained either by the fact that different kinds of simulations were taken into account (e.g., dark matter versus hydrodynamical runs with different treatments for the baryon physics) or by the fact that different volumes (e.g., within $R_{500}$ or $R_{200}$) were used to estimate the dynamical state. By considering the most external regions, indeed, more substructures that are still in the process of merging may be included, and the cluster result to be less virialised. For this reason, we decided to use the dynamical information provided by simulations to investigate the behaviour of the 12.7 \% most relaxed and 21.1 \% most disturbed objects of the sample (i. e., the 12.7\% of objects with the highest and the 21.1 \% of objects with the lowest values of $\chi$). These percentages represent the fraction of relaxed and disturbed objects of the CHEX-MATE sample identified in Subsect. \ref{final_classification}. 

\subsubsection{Morphological analysis of the simulated sample}\label{morpho_sim}
\begin{table*}[]
\centering
\renewcommand{\arraystretch}{1.3}
\caption{Comparison between the distributions of the observed and simulated morphological parameters.}\label{Tab:quartili}
\begin{tabular}{c|ccc|ccc|c}
\hline
\multirow{2}{*}{Parameter} & \multicolumn{3}{c|}{Observations} & \multicolumn{3}{c|}{Simulations} & \multirow{2}{*}{KS-test}  \\ \cline{2-7}
          & Median & 1$^{\text{st}}$ quartile & 3$^{\text{rd}}$ quartile & Median & 1$^{\text{st}}$ quartile & 3$^{\text{rd}}$ quartile & $p$-values  \\ \hline 
         c & 0.29       &   0.21         &     0.43       &  0.39 $\pm$ 0.02      & 0.28 $\pm$ 0.02 &   0.51 $\pm$ 0.02       &  0.003 $\pm$ 0.09         \\ 
         w &  0.011     &  0.005          &     0.024       & 0.019 $\pm$ 0.002       &  0.008 $\pm$ 0.001         &  0.041 $\pm$ 0.006     & 0.06 $\pm$ 0.09   \\ 
         P$_{\text{20}}$ &   2.6 $\cdot$ 10$^{-6}$     &     1.0 $\cdot$ 10$^{-6}$       &   7.5 $\cdot$ 10$^{-6}$         &   (3.2 $\pm$ 0.8)$\cdot$10$^{-6}$    &  (0.8 $\pm$ 0.2)$\cdot$10$^{-6}$           &    (16 $\pm$ 4)$\cdot$10$^{-6}$     & 0.2 $\pm$ 0.2  \\
         P$_{\text{30}}$  &   1.3 $\cdot$ 10$^{-7}$       &     0.3 $\cdot$ 10$^{-7}$         &  5.8 $\cdot$ 10$^{-7}$            & (2.1 $\pm$ 0.8)$\cdot$10$^{-7}$       &    (0.3 $\pm$ 0.2)$\cdot$10$^{-7}$        &    (16 $\pm$ 5)$\cdot$10$^{-7}$    & 0.13 $\pm$ 0.14   \\ 
         c$_{\text{no center}}$ & 0.32 & 0.26 & 0.38 & 0.44 $\pm$ 0.02 & 0.34 $\pm$ 0.02 & 0.54 $\pm$ 0.01 & (1 $\pm$ 2) $\cdot$ 10$^{-8}$ \\         \hline
   \end{tabular}
\tablefoot{ From left to right: name of the parameter, median and first and third quartile of the observed sample, median and first and third quartile of the simulated sample, results of the KS-test between the observed and simulated populations. The concentration in the last row is computed by excluding the central region (R $<$ 0.15 $R_{500}$) and using as inner aperture 0.3 $R_{500}$.}
\end{table*}

The X-ray images associated to each cluster are produced with Smac \citep{Dolag2005} without the inclusion of any background component and are filtered in the 0.7 - 1.2 keV energy band. The pixel size of each maps is 4 kpc fixed at all redshifts. In order to reproduce the  \xmm\ observations, we smoothed the simulated images with a Gaussian function of $\sigma$ = 6 arcsec, which represents the FWHM of the \xmm\ point spread function\footnote{https://xmm-tools.cosmos.esa.int/external~/xmm\_user\_support~/documentation/uhb/onaxisxraypsf.html}, and we binned them using the same scale of the observations (i.e., 1 pixel = 2.5 arcsec). Starting from these images we then computed the morphological parameters described in Sect. \ref{parametri} for each simulated clusters. 

To properly compare the distribution of the morphological parameters obtained from the simulated images with the observed ones, we randomly extracted from the 4692 X-ray observations of the 1564 simulated clusters, $10^4$ sub-samples, each consisting of 118 systems. Each sub-sample was build with the aim to reproduce the distribution in mass and redshift of the CHEX-MATE objects and avoiding the selection of the same cluster in more snapshot of the simulations. In particular, we considered only the simulated objects reproducing the properties of the Tier 1 and Tier 2, and we extract from each snapshot, $i$, a number of clusters corresponding to the number of the CHEX-MATE objects located at redshift $\frac{z_i-z_{i-1}}{2} < z < \frac{z_i+z_{i-1}}{2}$. By computing for each extraction the first, second and third quartile of the distribution of the morphological parameters, we obtained the mean values of these quantities and their associated uncertainties (namely their standard deviation). In Table \ref{Tab:quartili}, we report both a comparison of this result with the one arising from the observations and also the $p$-value of the Kolmogorov-Smirnov (KS) test derived to compare the distributions of the observed and simulated morphological parameters (also in this case the value and the uncertainties reported are respectively the mean and the standard deviation of the values obtained for each random-extracted sub-sample). We noticed a good agreement between the observed and simulated $P_{20}$, $P_{30}$, and $w$ (high $p$-values of the KS-Test, while $c$ shows a borderline behaviour which will better discussed in Sect \ref{concentration_discussion}. We also computed the statistical relative error associated to each parameter (defined as the ratio between the standard deviation and the mean of the the values obtained for each random-extracted sub-sample), finding values of 5\%, 10\%, 25\% and 38\% for $c$, $w$, $P_{20}$ and $P_{30}$ respectively.

For each random-extracted sub-sample we also evaluated the grade of correlation between these four parameters by estimating the Spearman coefficients. The results of this analysis are shown in the corner plot of Fig. \ref{morphology_oss}, right panel. In particular, in green are represented the couple of parameters for which the observed Spearman coefficient is within one standard deviation (1 $\sigma$) from the mean of the simulated Spearman coefficients, in yellow the Spearman coefficients between 1 and 3 $\sigma$ and in red the correlations over 3 $\sigma$. This comparison highlights that, all the couples show correlations included in the 3 $\sigma$ interval, with the exception of the $P_{20}$ -- $P_{30}$ whose trend does not reproduce the one arising from the observations. This behaviour could be related to the fact that in simulations the signal is strong up to $R_{500}$. 

As done for the observations, also in this case we combine the four parameters to build the parameter $M$ (see. Sect. for the definition). In Fig. \ref{fig:dyn_M} (right panel), we compare the distribution of this quantity with the distribution of the dynamical indicator $\chi$. As it is possible to observe, a correlation between the two quantities is present (Spearman coefficient, $r$=0.5). Furthermore it is possible to observe that the majority of the objects classified as relaxed (or disturbed) with the definition presented in Subsect. \ref{simulated_sample} are characterised by $M<0$ (or $M>0$) which is in agreement with the result obtained for the CHEX-MATE sample.

\begin{table}
\centering
\renewcommand{\arraystretch}{1.3}
\caption{Rank of the strongest correlations measured for the simulated sample.}
\begin{tabular}{@{}ccc@{}}
\toprule
Rank & Parameters & Spearman Coefficient  \\ \midrule
1) & P$_{\text{20}}$ -- P$_{\text{30}}$  & 0.85 $\pm$ 0.03  \\
2) &  P$_{\text{20}}$ -- w &  0.76 $\pm$ 0.04\\
3) &  P$_{\text{30}}$ -- w &  0.74 $\pm$ 0.04 \\
4) & w -- c & -0.67 $\pm$ 0.05  \\
5)&  P$_{\text{20}}$ -- c & -0.63 $\pm$ 0.06 \\
6) & P$_{\text{30}}$ -- c & -0.59 $\pm$ 0.06 \\ \bottomrule
\end{tabular}
 \end{table}

\subsubsection{Discussion on the concentration}\label{concentration_discussion}

\begin{figure}
    \centering
    \includegraphics[scale=0.45]{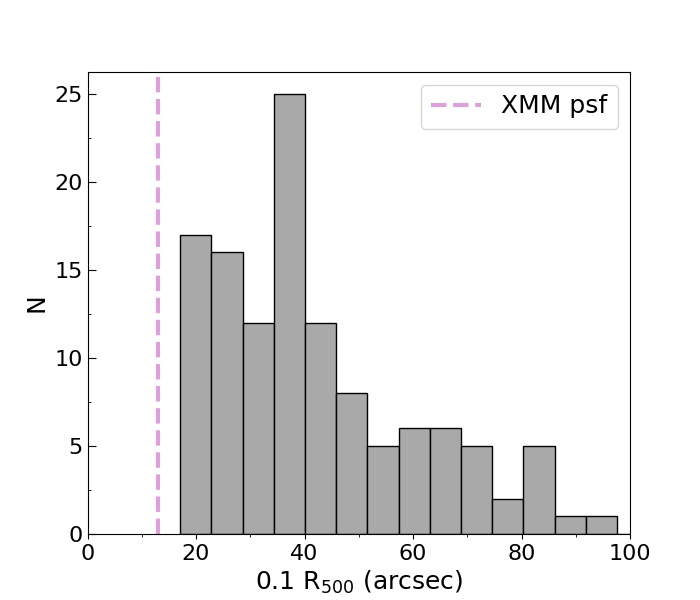}
    \caption{Comparison between the dimension of the inner aperture used for the computation of $c$ (i.e., $0.1 R_{500}$ and the dimension of the \xmm\ PSF (violet dashed line).}
    \label{fig:psf}
\end{figure}

As presented in Subsect.~\ref{morpho_sim}, some discrepancies are present between the observed and simulated distributions of the concentration. To investigate the nature of this differences, we first checked the validity of our procedure by comparing our result with the concentration estimated from surface brightness profiles by Bartalucci et al. 2022 (in prep.). In this latter analysis, the SB profiles were centred on the X-ray peak, then subtracted of the background and corrected for the vignetting. The concentration is then computed within two apertures of radii 0.1 $R_{500}$ and $R_{500}$, using the relation:
\begin{equation}
   C_{SB}= \frac{\int_{0}^{0.1 R_{500}} S_X(r)r dr}{\int_{0}^{R_{500}} S_X(r)r dr}.
\end{equation}
Eventually, the PSF is taken into account by using the King function \citep[e.g.,][]{Read2011}:
\begin{equation}
B(r)=\frac{A}{[1+(r/R_0)^2]^{\alpha}}    
\end{equation}
where $R_0=0.088$ arcmin is the core radius and $\alpha$ = 1.59 is the index. 
To be consistent with this analysis based on the profiles, we recomputed the values of the concentration using the same apertures. The results of this comparison is shown in Table \ref{Tab:comparison_Bart}. 
It is possible to observe that a good agreement is present between the two procedures, with the medians and the first and third quartiles being very similar in both cases. The lack of an influence of the PSF on the estimation of $c$ (which is actually observed in literature) is probably related to the choice of the radii in which $c$ is computed. As presented in Fig. \ref{fig:psf} indeed, 0.1 $R_{500}$ is larger than the dimension of the on-axis PSF for all the CHEX-MATE clusters, and consequently the photons located in the centre are not spread across larger regions. From this check we verified that the origin of the discrepancy between the observed and simulated sample is not related to the method adopted for the estimation of $c$.

We then focused on the simulated sample and we noticed that many systems with high $c$ values show a ring emission in their central region (i.e., $r<40$ kpc; for an example, see Fig.~\ref{fig:anello} in the Appendix~\ref{ring}). This effect is related to the presence of particles with high SPH density in the central regions of the simulated clusters, due to the action of the isotropic feedback from AGNs. In these simulations indeed, the AGN feedback is modelled as a single bubble which expands and compresses gas. Therefore, we re-computed $c$ for both the simulated and observed sample excluding the central region of the cluster (i.e. r < 0.15 R$_{500}$) and adopting as inner and outer apertures $0.3 \cdot R_{500}$ and $R_{500}$.
The results obtained are presented in Table \ref{Tab:quartili} (last raw). It is possible to observe, that a difference between the observed and simulated distributions is still present. Therefore, we concluded that the discrepancies between the CHEX-MATE X-ray images and simulated maps are not limited to the description of the core. A more accurate analysis of the simulated and observed SB profiles is necessary to better investigate the behaviour of the $c$ distributions.
\begin{table}
\centering
\renewcommand{\arraystretch}{1.3}
\caption{Comparison between the distribution of the concentration.} \label{Tab:comparison_Bart}
\begin{tabular}{@{}ccc@{}}
\toprule
                           & SB profiles & images \\ \midrule
\multicolumn{1}{c}{median} &   0.24    &     0.23    \\
1$^{st}$ quartile                 &   0.16    &     0.15     \\
3$^{rd}$ quartile                 &   0.37    &     0.35     \\ \bottomrule
\end{tabular}
\tablefoot{The values of the second and third column were computed from surface brightness profiles (see Bartalucci et al. in prep.) and from images, respectively. The apertures adopted for this comparison are 0.1 - 0.5 $R_{500}$.}
\end{table}

\subsubsection{Effects of the orientation}\label{orientation}
\begin{figure*}
    \hspace{-2cm}
    \includegraphics[scale=0.48]{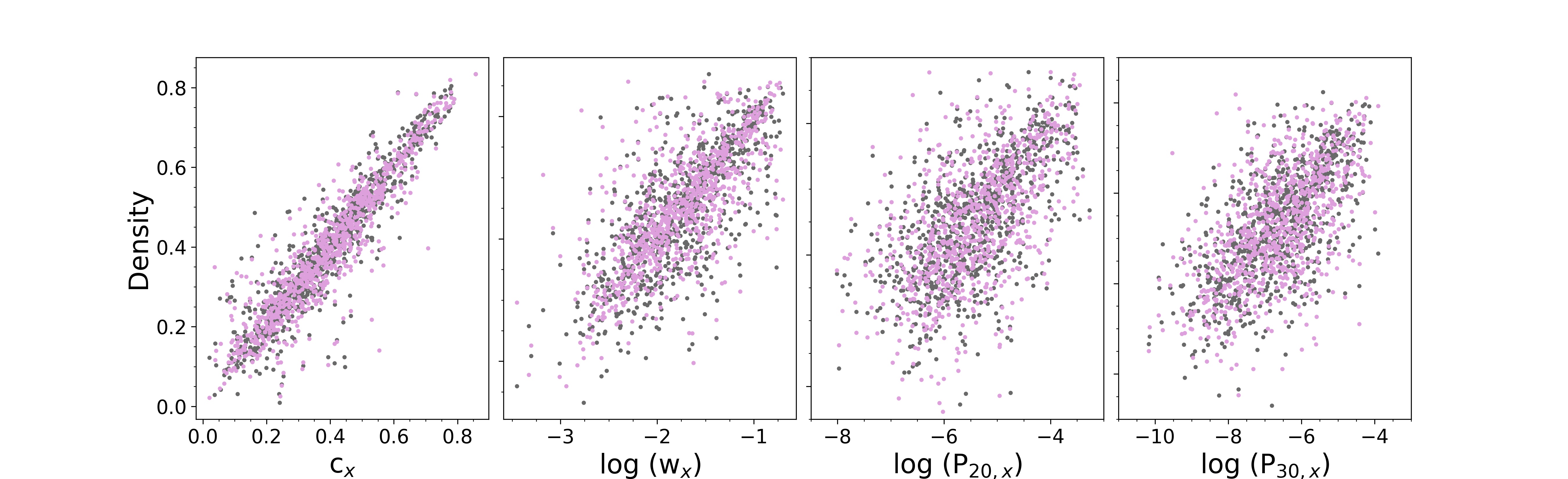}
   \caption{Comparison between the parameters computed along the $x$ direction (x-axis in the plot) and the $y$ ad $z$ directions (grey and violet respectively).}
    \label{fig:xyz}
    \end{figure*}

Finally, simulations are useful to understand how the orientation of a cluster can influence the estimation of the morphological parameters and consequently the classification of its dynamical state. For each of the 1564 simulated cluster we had at our disposal three images related to three different orientations of the object (namely, the orientations along the $x$, $y$, or $z$ axes\footnote{$x$,$y$ and $z$ are aligned with the cosmological box coordinates and are not correlated to the orientation of the major axis of the simulated clusters. Therefore the choice of the $x$ coordinate instead of the $y$ or $z$ does not impact the result.}). Therefore, we computed the morphological parameters for each orientation and we estimated the dispersion $\Delta$ between the values obtained for the $x$ projection and the values estimated along the other projections following the relation:
\begin{equation}
\Delta = |\log_{10}(\mathcal{P}_x)-\log_{10}(\mathcal{P}_{y,z})|
\end{equation}
with $\mathcal{P}$ representing a general parameter and the subscript representing the orientation considered ($x$, $y$ and $z$). The mean distribution of the dispersion along the $y$ and $z$ axis is shown in Figure \ref{fig:xyz}, first row. The median values of $\Delta$ obtained are 0.03, 0.17, 0.45 and 0.56 for $c$, $w$, $P_{20}$, and $P_{30}$, respectively.

\subsection{Robustness and efficiency of the parameters} \label{robustness}
Using the median value of the parameter $\Delta$ obtained from the investigation of the effects of the exposure time see Sect. \ref{exposure}), radius (see Sect. \ref{radius}) and orientation (see Sect. \ref{orientation}) on the estimates of the morphological parameters, we computed the associated relative error:
\begin{equation}
    \epsilon = 10^{\Delta}-1
\end{equation}
The results obtained are reported in Table \ref{percentage_error}. We noticed that the concentration is the most stable parameter, and it is also the indicator that shows the lowest uncertainties related to orientations effects and to the exposure time of the observations. For what concerns the other parameters, it may seems that the orientation is the major source of uncertainty, especially for the power ratios, which show percentage relative error of 180\% and 263\% for $P_{20}$ and $P_{30}$, respectively. However, we have to take into account the fact that these two parameters cover a large range of values. We found indeed, that for more than 93 \% of the clusters the variation of the power ratios with the orientation is less than the 10\% of the total observed range of values. This results is similar to the one obtained by \cite{Jeltema2008}.

The uncertainties related to the estimation of $w$, $P_{20}$ and $P_{30}$ increase with the decreasing of the exposure time. High values of the relative percentage errors are indeed obtained using images with $t_{{\mathrm{exp}}}$ equal to 5 ks. Given these results, we concluded that under the conditions of our analysis (i.e. high quality observations and low uncertainties on the $R_{500}$ estimation), our four parameters could be considered stable.

\begin{table}[]
\centering
\renewcommand{\arraystretch}{1.3}
\caption{Percentage relative error of the four morphological parameters obtained from the analysis of the systematics.}\label{percentage_error}
\begin{tabular}{lccccc}
\hline 
                    & \begin{tabular}[c]{@{}c@{}} $t_{{\mathrm{exp, 5ks}}}$ \\ (\%)\end{tabular} & \begin{tabular}[c]{@{}c@{}} $t_{{\mathrm{exp, 50 \%}}}$ \\ (\%)\end{tabular} & \begin{tabular}[c]{@{}c@{}} $R_{500 + 5\%}$ \\ (\%)\end{tabular} & \begin{tabular}[c]{@{}c@{}} $R_{500 - 5\%}$\\ (\%)\end{tabular} & \begin{tabular}[c]{@{}c@{}} orientation \\ (\%)\end{tabular} \\ \hline
$\epsilon_c$       &  0.93     &    0.70 & 4.7  & 4.7   & 7.2     \\
$\epsilon_w$ & 15         & 7.2  & 4.7   & 4.7  & 48 \\
$\epsilon_{P_{20}}$ & 26    & 12 &      17   &   15 & 180     \\
$\epsilon_{P_{30}}$ & 58 & 32   & 32 & 38 & 263 \\ \hline
\end{tabular}
\tablefoot{The results reported arise from the test related to: the exposure time of the images (first and second columns, see Sect. \ref{exposure}), the radius of the region considered (third and fourth columns, see Sect. \ref{radius}), and the orientation of the cluster (fifth column, see Sect. \ref{orientation}).}
\end{table}

Eventually, for each morphological parameter we assess the thresholds values above or below which the relaxed and disturbed systems (defined from the final classification, see Subsect. \ref{final_classification}) lie. We found that the relaxed population is characterised by values of $c >0.49$, $w<0.006$, $P_{20}<1.0\cdot10^{-6}$, $P_{30}<0.4\cdot10^{-7}$, while the disturbed population shows values of $c<0.19$, $w>0.01$, $P_{20}>4.0\cdot10^{-6}$ and $P_{30}>0.5\cdot 10^{-7}$.

\section{Summary and Conclusions}
In this paper, we present the morphological analysis performed on the X-ray images of the 118 clusters of the CHEX-MATE sample and we provide a classification of their dynamical state. To achieve this aim, we investigated the behaviour of four morphological parameters commonly used in literature to identify the  most relaxed and most disturbed systems. These parameters are: the concentration, $c$, the centroid shift, $w$ and the power ratios P$_{20}$ and P$_{30}$. To verify the ability of these indicators in reproducing the classification realised by trained eyes, we first realised a visual classification of the sample. In particular, seven astronomers where involved to inspect images by eye assigning to each object a grade from 0 (most relaxed) to 2 (most disturbed). By averaging these seven classifications, we found that 19 clusters are defined as most relaxed systems and 37 as most disturbed. All the other systems (62) are not clearly classifiable and are thus defined as mixed. Using this classification as reference for our analysis, we found that:
\begin{itemize}

\item The distributions of all the four morphological parameters do not show signs of bimodality (see Subsect. \ref{Sect:distributions}); 

\item the median (first-third quartile) of the distributions of the morphological parameters are respectively: 0.29 (0.21--0.43) for $c$; 0.011 (0.005--0.024) for $w$; 2.6 $\cdot$ 10$^{-6}$ (1.0--7.5 $\cdot$ 10$^{-6}$) for $P_{20}$ and 1.3 $\cdot$ 10$^{-7}$ (0.3--5.8 $\cdot$ 10$^{-7}$) for $P_{30}$ (see e.g., Table \ref{Tab:quartili});
\item the four morphological parameters are characterised by strong correlations (i.e. Spearman coefficient $r>0.5$, see Sect. \ref{correlazioni} and Fig. \ref{morphology_oss}, left panel), with the tails of their distributions which are populated by the most relaxed and most disturbed systems identified by the visual classification;
\item these four morphological parameters are robust enough to be not biased under the conditions of our analysis. They have proved to be not influenced by the exposure time of the observations and by uncertainties related to the definition of the region where they are computed. The systematic relative error due to the exposure time is: 0.7\%, 7.2 \%, 12\% and 32\% for $c$, $w$, $P_{20}$ and $P_{30}$, respectively, when the exposure time of the observations is halved (while is 0.9\%, 15\%, 26\% and 58\% for $c$, $w$, $P_{20}$ and $P_{30}$, when the exposure time of the observations is fixed to 5 ks; see Sect.~\ref{exposure}). For what concerns the  assumption of $R_{500}$, we found a systematic relative error of: $\sim$5\%, $\sim$5\%, $\sim$15\% and $\sim$32\% for $c$, $w$, $P_{20}$ and $P_{30}$, respectively (see Sect.~\ref{radius}); 
\item these values are on average lower than (or comparable with) the mean statistical relative errors of 5\%, 10\%, 25\% and 38\% for $c$, $w$, $P_{20}$ and $P_{30}$ respectively (see Subsect. \ref{morpho_sim}), confirming in particular how the concentration is less prone to systematic effects;
\item we combined the parameters in a single quantity, $M$, that is able to assign to each object of the sample a grade of relaxation. We noticed that also this new parameter do not show signs of bimodality. We then realised a continuous classification of the dynamical state of the sample based on the values of the $M$ parameter (see Sect. \ref{sect:Mparameter});
\item by comparing the visual classification and the classification based on the $M$ parameter, we finally classified as relaxed and as disturbed the objects for which the two classifications are in agreement. Using this criteria, we found that 15 systems (12.7 \%) are relaxed and 25 (21.1 \%) are disturbed (see Subsect. \ref{final_classification});
\item According to the classification of the previous bullet, we found that the relaxed population is characterised by values of $c >0.49$, $w<0.006$, $P_{20}<1.0\cdot10^{-6}$, $P_{30}<0.4\cdot10^{-7}$, while the disturbed population shows values of $c<0.19$, $w>0.01$, $P_{20}>4.0\cdot10^{-6}$ and $P_{30}>0.5\cdot 10^-7$ (see Subsect. \ref{robustness});
\item from the comparison of our analysis with previous works related to X-ray selected samples, we found that the CHEX-MATE objects tend to be more dynamically disturbed (i.e., high centroid shift and low concentration) than the X-ray selected samples, in agreement with what has been obtained by other recent studies (see Subsect. \ref{comparison_oss});
\item by repeating our analysis on a simulated sample provided by {\sc The Three Hundred} collaboration, we found that a good agreement is present for what concerns the level of correlation between the parameter pairs (see e.g., Fig. \ref{morphology_oss}, right panel). However, the comparison between the observed and simulated distributions of the morphological parameters highlighted the presence of a discrepancy: the simulated values of the concentration are on average higher than the observed ones (see Subsect. \ref{morpho_sim});
\item by investigating the behaviour of $c$, we found that simulations are characterised by a distribution with higher values due to the non-negligible presence of particles with high SPH density produced from the action of the isotropic thermal AGN feedback in the central regions of the simulated objects. This behaviour indicates that a more realistic and higher resolution feedback model is required (such as a circulating mechanical anisotropic AGN feedback, see Subsect. \ref{concentration_discussion}). However, the discrepancy between the observed and disturbed distribution remains even if the core region is masked when computing the morphological parameters. Therefore, further analysis are necessary, to investigate the properties of the observed and simulated SB profiles;
\item thanks to simulations we investigated the systematic relative error associated to the orientation of the cluster, finding values of 7\%, 48\%, 180\% and 263\% for $c$, $w$, $P_{20}$ and $P_{30}$ (see Subsect. \ref{morpho_sim}), which represent the dominant component in their error budget, as expected for quantities based on the projected distribution of X-ray counts in the plane of the sky.
\end{itemize}

This study of the X-ray morphological properties of the CHEX-MATE objects provides
the parameters that will be used in forthcoming analyses to assess the role of the dynamical state in the reconstruction and characterisation of their intrinsic physical quantities, from the thermodynamic profiles to their distribution in the scaling relations.

\section*{Acknowledgements}
SE, LL, IB, MR, FG, SG and SM acknowledge financial contribution from the contracts ASI-INAF Athena 2019-27-HH.0,
``Attivit\`a di Studio per la comunit\`a scientifica di Astrofisica delle Alte Energie e Fisica Astroparticellare''
(Accordo Attuativo ASI-INAF n. 2017-14-H.0), INAF mainstream project 1.05.01.86.10, and
funding from the European Union’s Horizon 2020 Programme under the AHEAD2020 project (grant agreement n. 871158). GWP acknowledges support from CNES, the French space angency.

\bibliographystyle{aa}
\bibliography{aanda}

\begin{appendix}

\section{Parameters values}\label{Appendix:parameters_values}
All the parameter values used in this paper and calculated
within R$_{\text{500}}$ are listed in Table \ref{Tab:final_results}. The objects marked with an asterisk are those for which the estimation of the morphological parameters could be may not be accurate. These systems are:
\begin{itemize}
    \item G048.10+57.16, for which the emission inside $R_{500}$ is not fully covered if we choose as centre the X-ray peak;
    \item G283.91+73.97, for which the characterisation of the background is complex, since it is located behind Virgo;
    \item G028.63+50.15, whose X-ray emission is very complex and may not be described accurately;
    \item G107.10+65.32 and G124.20-36.48, which show a disturbed morphology, due to the ongoing merger between two sub-structures of similar dimensions. Since a double X-ray peak is present, and since it is not possible to identify the principal cluster, the parameters could be influenced by the choice of the X-ray peak.  
\end{itemize}
\onecolumn

\begin{landscape}

\begin{ThreePartTable}

\begin{TableNotes}
\footnotesize
\item[\bfseries Note:] From left to right: name, second moment of the power ratio, $P_{20}$, third moment of the power ratio, $P_{30}$, concentration, $c$, centroid shift, $w$, asymmetry, $A$ (see Appendix \ref{Other_parameters} for definition), smoothness, $S$ (see Appendix \ref{Other_parameters} for definition), ellipticity, $\eta$, results from the visual classification (see Sect. \ref{visual classification} for more details), parameter $M$ (see Sect. \ref{sect:Mparameter} for more details), dynamical state based on the comparison between $M$ and the visual classification, rank of the dynamical state of the sample (from the most relaxed 1, to the most disturbed, 118). The asterisks identify the systems for which the estimation of the morphological parameters may not be accurate (see Appendix \ref{Appendix:parameters_values} for more details).
\end{TableNotes}

\begin{longtable}{llllllllllll}
\caption{Morphological parameters for the 118 CHEX-MATE clusters within R$_{500}$.}\label{Tab:final_results}\\ 
\hline \hline \smallskip
Name & P20 & P30 & c & w & A & S & $\eta$ & Visual & M & State & Rank \\
 & ($\times$ 10$^{-6}$) & ($\times$ 10$^{-7}$) & & ($\times$ 10$^{-1}$) & & & & & & &\\
\hline \hline
\endfirsthead
\caption{continued}\\
\hline\hline
Name & P20 (e-6) & P30 (e-7) & c & w (e-1) & A & S & $\eta$ & Visual & M & State & Rank\\
\hline
\endhead
\hline
\endfoot  
\insertTableNotes
\endlastfoot
G000.13+78.04 & 1.4$^{+0.8}_{-1.2 }$      &10  $^{+3   }_{-5   }$ & 0.20$^{+0.05}_{-0.05}$  & 0.32 $^{+0.10  }_{-0.12  }$ & 1.1$^{+0.3}_{-0.3}$ & 0.90$^{+0.17}_{-0.16}$   & 0.92$^{+0.07}_{-0.05}$ & 1.2  $\pm$ 0.4  & 0.59   & M  &   86    \\ 
G004.45-19.55 & 1.5$^{+0.3}_{-1.1 }$ 	  &3   $^{+1   }_{-3   }$ & 0.37$^{+0.09}_{-0.01}$  & 0.06 $^{+0.02  }_{-0.04  }$ & 0.4$^{+0.4}_{-0.3}$ & 0.27$^{+0.13}_{-0.11}$   & 0.97$^{+0.01}_{-0.03}$ & 0.8  $\pm$ 0.4  & -0.28 &  M  &   45  \\
G006.49+50.56 & 0.1$^{+0.4}_{-0.4 }$ 	  &0.04$^{+0.06}_{-0.01}$ & 0.53$^{+0.04}_{-0.04}$  & 0.020$^{+0.003 }_{-0.001 }$ & 0.9$^{+0.1}_{-0.1}$ & 0.77$^{+0.11}_{-0.10}$   & 0.97$^{+0.04}_{-0.05}$ & 0.2  $\pm$ 0.4  & -1.65  & R  &    3 \\ 
G008.31-64.74 & 33.0$^{+9}_{-12. }$  	  &5   $^{+1   }_{-3   }$ & 0.21$^{+0.02}_{-0.03}$  & 0.19 $^{+0.02  }_{-0.03  }$ & 1.1$^{+0.3}_{-0.2}$ & 0.89$^{+0.16}_{-0.15}$   & 0.65$^{+0.13}_{-0.11}$ & 1.7  $\pm$ 0.5  & 0.82  &  D  &  100  \\
G008.94-81.22 & 8.3 $^{+1.4}_{-1.9 }$	  &17  $^{+4   }_{-5   }$ & 0.19$^{+0.09}_{-0.01}$  & 0.53 $^{+0.04  }_{-0.05  }$ & 0.8$^{+0.1}_{-0.1}$ & 0.38$^{+0.09}_{-0.08}$   & 0.85$^{+0.04}_{-0.03}$ & 2.0  $\pm$ 0.0  & 1.06  &  D  &  108 \\ 
G021.10+33.24 & 0.10$^{+0.07}_{-0.09}$    &0.15$^{+0.12}_{-0.15}$ & 0.66$^{+0.01}_{-0.01}$  & 0.022$^{+0.003 }_{-0.010 }$ & 0.7$^{+0.2}_{-0.2}$ & 0.65$^{+0.04}_{-0.04}$   & 0.92$^{+0.08}_{-0.06}$ & 0.0  $\pm$ 0.0  & -1.59 &  R  &    4\\ 
G028.63+50.15* & 42$^{+29}_{-30. }$        &30  $^{+0   }_{-20  }$ & 0.19$^{+0.07}_{-0.08}$  & 1.2  $^{+0.5   }_{-0.6   }$ & 1.5$^{+0.4}_{-0.4}$ & 1.1 $^{+0.2 }_{-0.2 }$   & 0.75$^{+0.16}_{-0.15}$ & 2.0  $\pm$ 0.0  & 1.56  &  D  &  117 \\ 
G028.89+60.13 & 0.8$^{+0.2}_{-0.4 }$      &0.10$^{+0.09}_{-0.06}$ & 0.52$^{+0.08}_{-0.08}$  & 0.018$^{+0.011 }_{-0.011 }$ & 1.1$^{+0.2}_{-0.2}$ & 0.91$^{+0.10}_{-0.09}$   & 0.84$^{+0.10}_{-0.09}$ & 0.2  $\pm$ 0.4  & -1.25 &  R  &   11 \\ 
G031.93+78.71 & 0.09$^{+0.01}_{-0.06}$    &1.4 $^{+0.2 }_{-0.6 }$ & 0.31$^{+0.07}_{-0.07}$  & 0.148$^{+0.001 }_{-0.010 }$ & 1.3$^{+0.3}_{-0.3}$ & 1.05$^{+0.15}_{-0.15}$   & 0.95$^{+0.04}_{-0.02}$ & 1.0  $\pm$ 0.0  & -0.47 &  M  &   35 \\ 
G033.81+77.18 & 0.1$^{+0.2}_{-0.2 }$      &0.01$^{+0.01}_{-0.01}$ & 0.54$^{+0.02}_{-0.03}$  & 0.057$^{+0.005 }_{-0.010 }$ & 0.8$^{+0.1}_{-0.1}$ & 0.59$^{+0.08}_{-0.08}$   & 0.96$^{+0.03}_{-0.04}$ & 0.3  $\pm$ 0.5  & -1.58 &  R  &    5 \\ 
G040.03+74.95 & 3$^{+1}_{-1   }$          &1.0 $^{+0.6 }_{-0.9 }$ & 0.34$^{+0.01}_{-0.01}$  & 0.032$^{+0.002 }_{-0.010 }$ & 1.5$^{+0.4}_{-0.4}$ & 1.3 $^{+0.3 }_{-0.3 }$   & 0.86$^{+0.07}_{-0.06}$ & 1.0  $\pm$ 0.6  & -0.41 &  M  &   41 \\ 
G040.58+77.12 & 7.9$^{+4.0}_{-5.0 }$      &11  $^{+6   }_{-7   }$ & 0.30$^{+0.01}_{-0.01}$  & 0.24 $^{+0.11  }_{-0.13  }$ & 1.4$^{+0.4}_{-0.4}$ & 1.1 $^{+0.2 }_{-0.2 }$   & 0.7 $^{+0.2 }_{-0.2 }$ & 1.2  $\pm$ 0.4  & 0.56  &  M  &   85 \\ 
G041.45+29.10 & 1.6$^{+0.4}_{-0.8 }$      &0.7 $^{+0.3 }_{-0.2 }$ & 0.16$^{+0.03}_{-0.03}$  & 0.31 $^{+0.05  }_{-0.07  }$ & 1.0$^{+0.2}_{-0.2}$ & 0.67$^{+0.12}_{-0.12}$   & 0.95$^{+0.03}_{-0.02}$ & 2.0  $\pm$ 0.0  & 0.40  &  M  &   78  \\
G042.81+56.61 & 2.7$^{+0.3}_{-0.6 }$      &0.53$^{+0.07}_{-0.10}$ & 0.29$^{+0.03}_{-0.03}$  & 0.15 $^{+0.01  }_{-0.02  }$ & 1.2$^{+0.2}_{-0.2}$ & 0.99$^{+0.14}_{-0.14}$   & 0.85$^{+0.05}_{-0.04}$ & 1.0  $\pm$ 0.0  & -0.05 &  M  &   56 \\ 
G044.20+48.66 & 1.8$^{+0.6}_{-0.6 }$      &0.60$^{+0.07}_{-0.01}$ & 0.43$^{+0.02}_{-0.02}$  & 0.108$^{+0.001 }_{-0.003 }$ & 0.8$^{+0.1}_{-0.1}$ & 0.56$^{+0.09}_{-0.08}$   & 0.83$^{+0.06}_{-0.05}$ & 1.0  $\pm$ 0.0  & -0.39 &  M  &   42 \\ 
G044.77-51.30 & 1.1$^{+0.5}_{-0.4 }$      &3   $^{+1   }_{-2   }$ & 0.34$^{+0.06}_{-0.07}$  & 0.06 $^{+0.02  }_{-0.01  }$ & 0.3$^{+0.3}_{-0.3}$ & 0.22$^{+0.13}_{-0.11}$   & 0.89$^{+0.07}_{-0.04}$ & 0.8  $\pm$ 0.4  & -0.28 &  M  &   46 \\ 
G046.10+27.18 & 10.$^{+4.}_{-5.  }$       &3   $^{+1   }_{-3   }$ & 0.15$^{+0.02}_{-0.03}$  & 0.22 $^{+0.05  }_{-0.08  }$ & 0.4$^{+0.3}_{-0.3}$ & 0.36$^{+0.15}_{-0.14}$   & 0.81$^{+0.10}_{-0.08}$ & 2.0  $\pm$ 0.0  & 0.80  &  D  &   99 \\ 
G046.88+56.48 & 22$^{+12}_{-13.}$        &60  $^{+30  }_{-40  }$ & 0.14$^{+0.04}_{-0.04}$  & 0.33 $^{+0.17  }_{-0.18  }$ & 1.3$^{+0.3}_{-0.3}$ & 1.05$^{+0.16}_{-0.16}$   & 0.74$^{+0.14}_{-0.13}$ & 2.0  $\pm$ 0.0  & 1.41  &  D  &  111 \\ 
G048.10+57.16* & 12.$^{+5.}_{-5.  }$       &5.9 $^{+1.1 }_{-1.7 }$ & 0.12$^{+0.02}_{-0.02}$  & 0.225$^{+0.004 }_{-0.010 }$ & 1.2$^{+0.2}_{-0.2}$ & 0.92$^{+0.13}_{-0.13}$   & 0.76$^{+0.09}_{-0.09}$ & 2.0  $\pm$ 0.0  & 1.03  &  D  &  106 \\ 
G049.22+30.87 & 1.2$^{+0.8}_{-0.5 }$      &0.1 $^{+0.7 }_{-0.4 }$ & 0.62$^{+0.08}_{-0.09}$  & 0.010$^{+0.003 }_{-0.002 }$ & 0.9$^{+0.2}_{-0.1}$ & 0.61$^{+0.07}_{-0.07}$   & 0.75$^{+0.12}_{-0.10}$ & 0.2  $\pm$ 0.4  & -1.43 &  R  &    7 \\  
G049.32+44.37 & 4$^{+2}_{-3   }$          &0.4 $^{+0.6 }_{-0.1 }$ & 0.28$^{+0.06}_{-0.07}$  & 0.22 $^{+0.05  }_{-0.07  }$ & 1.5$^{+0.3}_{-0.3}$ & 1.3 $^{+0.2 }_{-0.2 }$   & 0.79$^{+0.14}_{-0.11}$ & 1.0  $\pm$ 0.0  & 0.08  &  M  &   59 \\ 
G050.40+31.17 & 1.6$^{+0.7}_{-0.9 }$      &0.14$^{+0.16}_{-0.04}$ & 0.36$^{+0.05}_{-0.05}$  & 0.11 $^{+0.02  }_{-0.03  }$ & 0.7$^{+0.2}_{-0.2}$ & 0.57$^{+0.11}_{-0.10}$   & 0.85$^{+0.09}_{-0.08}$ & 1.0  $\pm$ 0.0  & -0.48 &  M  &   33 \\ 
G053.53+59.52 & 4$^{+7}_{-10  }$          &120 $^{+30  }_{-40  }$ & 0.22$^{+0.03}_{-0.04}$  & 0.31 $^{+0.04  }_{-0.06  }$ & 1.5$^{+0.3}_{-0.3}$ & 1.20$^{+0.14}_{-0.13}$   & 0.61$^{+0.15}_{-0.13}$ & 1.2  $\pm$ 0.4  & 0.98  &  M  &   92 \\ 
G055.59+31.85 & 2.0$^{+0.1}_{-0.5 }$      &1.3 $^{+0.6 }_{-0.9 }$ & 0.46$^{+0.05}_{-0.05}$  & 0.06 $^{+0.02  }_{-0.01  }$ & 0.7$^{+0.2}_{-0.2}$ & 0.56$^{+0.10}_{-0.09}$   & 0.82$^{+0.07}_{-0.06}$ & 0.5  $\pm$ 0.5  & -0.46 &  M  &   39 \\ 
G056.77+36.32 & 0.9$^{+0.2}_{-0.3 }$      &1.0 $^{+0.4 }_{-0.6 }$ & 0.49$^{+0.06}_{-0.06}$  & 0.029$^{+0.004 }_{-0.001 }$ & 0.9$^{+0.2}_{-0.2}$ & 0.69$^{+0.09}_{-0.09}$   & 0.85$^{+0.08}_{-0.07}$ & 0.3  $\pm$ 0.5  & -0.81 &  M  &   21 \\ 
G056.93-55.08 & 8$^{+2}_{-3   }$          &1.9 $^{+0.3 }_{-0.4 }$ & 0.17$^{+0.01}_{-0.01}$  & 0.25 $^{+0.02  }_{-0.03  }$ & 0.5$^{+0.1}_{-0.1}$ & 0.26$^{+0.09}_{-0.08}$   & 0.80$^{+0.07}_{-0.06}$ & 1.5  $\pm$ 0.5  & 0.67  &  D  &   96 \\ 
G057.25-45.34 & 1.2$^{+0.1}_{-0.3 }$      &1.0 $^{+0.2 }_{-0.4 }$ & 0.46$^{+0.03}_{-0.03}$  & 0.080$^{+0.005 }_{-0.018 }$ & 0.6$^{+0.2}_{-0.2}$ & 0.40$^{+0.11}_{-0.10}$   & 0.86$^{+0.04}_{-0.03}$ & 0.7  $\pm$ 0.5  & -0.49 &  M  &   31 \\  
G057.61+34.93 & 2.3$^{+0.7}_{-1.0 }$      &9   $^{+3   }_{-4   }$ & 0.16$^{+0.03}_{-0.04}$  & 0.10 $^{+0.05  }_{-0.06  }$ & 1.4$^{+0.3}_{-0.3}$ & 1.15$^{+0.19}_{-0.19}$   & 0.91$^{+0.04}_{-0.03}$ & 1.3  $\pm$ 0.5  & 0.49  &  M  &   82 \\  
G057.78+52.32 & 29 $^{+18}_{-19  }$       &60  $^{+30. }_{-40  }$ & 0.25$^{+0.08}_{-0.08}$  & 0.26 $^{+0.14  }_{-0.15  }$ & 1.5$^{+0.3}_{-0.3}$ & 1.2 $^{+0.2 }_{-0.2 }$   & 0.70$^{+0.18}_{-0.16}$ & 1.3  $\pm$ 0.5  & 1.08  &  M  &   93 \\  
G057.92+27.64 & 2.6$^{+1.5}_{-1.7 }$      &2.8 $^{+0.4 }_{-0.2 }$ & 0.58$^{+0.09}_{-0.01}$  & 0.05 $^{+0.02  }_{-0.02  }$ & 1.0$^{+0.2}_{-0.2}$ & 0.83$^{+0.11}_{-0.10}$   & 0.74$^{+0.18}_{-0.17}$ & 1.0  $\pm$ 0.0  & -0.49 &  M  &   32 \\  
G062.46-21.35 & 0.1$^{+0.2}_{-0.1 }$      &0.4 $^{+1.3 }_{-0.5 }$ & 0.52$^{+0.01}_{-0.01}$  & 0.06 $^{+0.01  }_{-0.01  }$ & 0.9$^{+0.3}_{-0.3}$ & 0.73$^{+0.13}_{-0.12}$   & 0.93$^{+0.04}_{-0.03}$ & 0.0  $\pm$ 0.0  & -1.10 &  R  &   13 \\  
G066.41+27.03 & 38$^{+14}_{-17  }$        &0.4 $^{+0.9 }_{-0.1 }$ & 0.15$^{+0.01}_{-0.02}$  & 0.21 $^{+0.04  }_{-0.06  }$ & 0.3$^{+0.2}_{-0.2}$ & 0.20$^{+0.11}_{-0.10}$   & 0.68$^{+0.12}_{-0.11}$ & 1.8  $\pm$ 0.4  & 0.74  &  D  &   98 \\ 
G066.68+68.44 & 0.47$^{+0.14}_{-0.04}$    &0.25$^{+0.11}_{-0.05}$ & 0.47$^{+0.05}_{-0.06}$  & 0.033$^{+0.001 }_{-0.010 }$ & 0.6$^{+0.2}_{-0.2}$ & 0.53$^{+0.09}_{-0.09}$   & 0.88$^{+0.05}_{-0.04}$ & 0.2  $\pm$ 0.4  & -1.02 &  R  &   15 \\ 
G067.17+67.46 & 0.16$^{+0.05}_{-0.10}$    &0.11$^{+0.04}_{-0.10}$ & 0.48$^{+0.04}_{-0.05}$  & 0.033$^{+0.003 }_{-0.001 }$ & 0.8$^{+0.2}_{-0.2}$ & 0.61$^{+0.10}_{-0.09}$   & 0.93$^{+0.04}_{-0.03}$ & 1.0  $\pm$ 0.0  & -1.29 &  M  &   16 \\ 
G067.52+34.75 & 0.7$^{+0.1}_{-0.2 }$      &0.5 $^{+0.1 }_{-0.4 }$ & 0.52$^{+0.09}_{-0.01}$  & 0.06 $^{+0.01  }_{-0.02  }$ & 0.9$^{+0.2}_{-0.2}$ & 0.75$^{+0.12}_{-0.11}$   & 0.88$^{+0.07}_{-0.04}$ & 0.5  $\pm$ 0.5  & -0.79 &  M  &   22 \\  
G068.22+15.18 & 1.8$^{+1.1}_{-1.3 }$      &0.5 $^{+0.1 }_{-0.3 }$ & 0.20$^{+0.05}_{-0.05}$  & 0.19 $^{+0.04  }_{-0.05  }$ & 1.5$^{+0.3}_{-0.3}$ & 1.3 $^{+0.2 }_{-0.2 }$   & 0.91$^{+0.06}_{-0.05}$ & 0.8  $\pm$ 0.4  & 0.14  &  M  &   63 \\  
G071.63+29.78 & 10$^{+5}_{-6   }$         &3   $^{+1   }_{-3   }$ & 0.13$^{+0.03}_{-0.04}$  & 0.29 $^{+0.08  }_{-0.10  }$ & 1.2$^{+0.3}_{-0.3}$ & 1.05$^{+0.19}_{-0.18}$   & 0.90$^{+0.05}_{-0.02}$ & 1.8  $\pm$ 0.4  & 0.94  &  D  &  103 \\ 
G072.62+41.46 & 0.9$^{+0.1}_{-0.4 }$      &0.9 $^{+0.1 }_{-0.4 }$ & 0.32$^{+0.04}_{-0.04}$  & 0.05 $^{+0.01  }_{-0.02  }$ & 1.0$^{+0.3}_{-0.3}$ & 0.81$^{+0.13}_{-0.13}$   & 0.86$^{+0.07}_{-0.05}$ & 1.0  $\pm$ 0.0  & -0.46 &  M  &   38 \\ 
G073.97-27.82 & 1.551$^{+0.478}_{-0.007}$ &0.3 $^{+0.1 }_{-0.2 }$ & 0.44$^{+0.05}_{-0.05}$  & 0.11 $^{+0.02  }_{-0.03  }$ & 1.0$^{+0.2}_{-0.2}$ & 0.75$^{+0.12}_{-0.11}$   & 0.83$^{+0.06}_{-0.04}$ & 0.8  $\pm$ 0.4  & -0.50 &  M  &   30 \\ 
G075.71+13.51 & 0.165$^{+0.010}_{-0.013}$ &0.06$^{+0.01}_{-0.01}$ & 0.33$^{+0.03}_{-0.03}$  & 0.26 $^{+0.02  }_{-0.02  }$ & 1.1$^{+0.2}_{-0.2}$ & 0.84$^{+0.11}_{-0.11}$   & 0.97$^{+0.00}_{-0.00}$ & 1.2  $\pm$ 0.4  & -0.66 &  M  &   26 \\ 
G077.90-26.63 & 0.3$^{+0.2}_{-0.1  }$     &0.1 $^{+0.3 }_{-0.1 }$ & 0.40$^{+0.06}_{-0.07}$  & 0.046$^{+0.002 }_{-0.010 }$ & 0.9$^{+0.2}_{-0.2}$ & 0.75$^{+0.12}_{-0.12}$   & 0.92$^{+0.03}_{-0.02}$ & 0.8  $\pm$ 0.4  & -1.03 &  M  &   19 \\ 
G080.16+57.65 & 23$^{+13}_{-15   }$       &22. $^{+15  }_{-17  }$ & 0.15$^{+0.06}_{-0.06}$  & 0.6  $^{+0.2   }_{-0.2   }$ & 1.5$^{+0.4}_{-0.4}$ & 1.2 $^{+0.3 }_{-0.3 }$   & 0.77$^{+0.13}_{-0.12}$ & 1.8  $\pm$ 0.4  & 1.40  &  D  &  110 \\  
G080.37+14.64 & 0.6$^{+0.4}_{-0.5  }$     &0.01$^{+0.02}_{-0.04}$ & 0.25$^{+0.07}_{-0.08}$  & 0.12 $^{+0.05  }_{-0.07  }$ & 1.2$^{+0.3}_{-0.3}$ & 1.00$^{+0.18}_{-0.17}$   & 0.91$^{+0.07}_{-0.05}$ & 1.2  $\pm$ 0.4  & -0.72 &  M  &   24 \\ 
G080.41-33.24 & 0.7$^{+0.7}_{-0.4  }$     &0.6 $^{+0.4 }_{-0.1 }$ & 0.36$^{+0.07}_{-0.07}$  & 0.28 $^{+0.03  }_{-0.03  }$ & 1.0$^{+0.2}_{-0.2}$ & 0.68$^{+0.10}_{-0.10}$   & 0.92$^{+0.02}_{-0.01}$ & 1.8  $\pm$ 0.4  & -0.20 &  M  &   50 \\ 
G083.29-31.03 & 4$^{+1}_{-2    }$         &8   $^{+2   }_{-4   }$ & 0.27$^{+0.03}_{-0.03}$  & 0.14 $^{+0.03  }_{-0.05  }$ & 0.6$^{+0.2}_{-0.2}$ & 0.37$^{+0.12}_{-0.11}$   & 0.86$^{+0.07}_{-0.05}$ & 1.3  $\pm$ 0.5  & 0.36  &  M  &   75 \\ 
G083.86+85.09 & 0.9$^{+0.1}_{-0.4  }$     &1.1 $^{+0.3 }_{-0.8 }$ & 0.31$^{+0.05}_{-0.05}$  & 0.10 $^{+0.01  }_{-0.02  }$ & 0.8$^{+0.3}_{-0.2}$ & 0.67$^{+0.14}_{-0.13}$   & 0.90$^{+0.05}_{-0.04}$ & 1.2  $\pm$ 0.4  & -0.26 &  M  &   48 \\ 
G085.98+26.69 & 1.2$^{+0.4}_{-0.8  }$     &2.9 $^{+0.2 }_{-1.6 }$ & 0.17$^{+0.05}_{-0.06}$  & 0.31 $^{+0.11  }_{-0.13  }$ & 1.0$^{+0.3}_{-0.3}$ & 0.81$^{+0.17}_{-0.16}$   & 0.96$^{+0.04}_{-0.01}$ & 1.5  $\pm$ 0.5  & 0.50  &  D  &   94 \\ 
G087.03-57.37 & 2.6$^{+0.6}_{-1.4  }$     &0.5 $^{+0.7 }_{-0.2 }$ & 0.26$^{+0.03}_{-0.04}$  & 0.31 $^{+0.03  }_{-0.05  }$ & 1.1$^{+0.3}_{-0.3}$ & 0.94$^{+0.14}_{-0.13}$   & 0.88$^{+0.06}_{-0.04}$ & 1.5  $\pm$ 0.5  & 0.17  &  M  &   64 \\ 
G092.71+73.46 & 22$^{+6}_{-8	}$        &19  $^{+6   }_{-9   }$ & 0.29$^{+0.05}_{-0.06}$  & 0.049$^{+0.0010}_{-0.018 }$ & 1.2$^{+0.3}_{-0.3}$ & 1.01$^{+0.13}_{-0.12}$   & 0.64$^{+0.15}_{-0.13}$ & 1.3  $\pm$ 0.5  & 0.42  &  M  &   79 \\ 
G094.69+26.36 & 2.5$^{+0.9}_{-1.7  }$     &1.5 $^{+0.2 }_{-1.2 }$ & 0.22$^{+0.07}_{-0.07}$  & 0.14 $^{+0.07  }_{-0.10  }$ & 1.1$^{+0.4}_{-0.4}$ & 1.00$^{+0.18}_{-0.17}$   & 0.88$^{+0.08}_{-0.06}$ & 1.2  $\pm$ 0.4  & 0.20  &  M  &   67 \\ 
G098.44+56.59 & 5$^{+2}_{-2    }$         &4   $^{+3   }_{-4   }$ & 0.20$^{+0.06}_{-0.06}$  & 0.26 $^{+0.07  }_{-0.09  }$ & 1.1$^{+0.3}_{-0.3}$ & 0.84$^{+0.16}_{-0.16}$   & 0.83$^{+0.10}_{-0.09}$ & 1.2  $\pm$ 0.4  & 0.62  &  M  &   87 \\ 
G099.48+55.60 & 21$^{+16}_{-17   }$       &19  $^{+11  }_{-14  }$ & 0.14$^{+0.05}_{-0.05}$  & 0.6  $^{+0.2   }_{-0.2   }$ & 1.3$^{+0.4}_{-0.4}$ & 1.1 $^{+0.2 }_{-0.2 }$   & 0.80$^{+0.15}_{-0.14}$ & 2.0  $\pm$ 0.0  & 1.41  &  D  &  112 \\  
G105.55+77.21 & 4$^{+2}_{-3    }$         &0.6 $^{+0.1 }_{-0.4 }$ & 0.27$^{+0.01}_{-0.01}$  & 0.18 $^{+0.09  }_{-0.11  }$ & 1.4$^{+0.4}_{-0.4}$ & 1.2 $^{+0.3 }_{-0.3 }$   & 0.80$^{+0.16}_{-0.15}$ & 1.2  $\pm$ 0.4  & 0.10  &  M  &   60 \\ 
G106.87-83.23 & 0.1$^{+0.2}_{-0.1  }$     &1.2 $^{+0.4 }_{-1.0 }$ & 0.34$^{+0.05}_{-0.05}$  & 0.09 $^{+0.01  }_{-0.03  }$ & 0.7$^{+0.3}_{-0.2}$ & 0.59$^{+0.14}_{-0.12}$   & 0.96$^{+0.02}_{-0.01}$ & 0.8  $\pm$ 0.4  & -0.65 &  M  &   27 \\ 
G107.10+65.32* & 4$^{+1}_{-2    }$         &2.1 $^{+0.6 }_{-0.6 }$ & 0.19$^{+0.03}_{-0.04}$  & 0.45 $^{+0.08  }_{-0.10  }$ & 0.9$^{+0.3}_{-0.2}$ & 0.59$^{+0.13}_{-0.12}$   & 0.87$^{+0.08}_{-0.06}$ & 2.0  $\pm$ 0.0  & 0.66  &  D  &   95 \\ 
G111.61-45.71 & 2.7$^{+0.2}_{-0.9  }$     &0.6 $^{+0.3 }_{-0.5 }$ & 0.23$^{+0.02}_{-0.03}$  & 0.062$^{+0.005 }_{-0.016 }$ & 0.2$^{+0.3}_{-0.2}$ & 0.21$^{+0.14}_{-0.13}$   & 0.88$^{+0.05}_{-0.02}$ & 1.2  $\pm$ 0.4  & -0.12 &  M  &   52 \\ 
G111.75+70.37 & 30$^{+11}_{-14   }$       &15  $^{+7   }_{-9   }$ & 0.13$^{+0.03}_{-0.03}$  & 0.6  $^{+0.2   }_{-0.2   }$ & 1.2$^{+0.3}_{-0.3}$ & 0.94$^{+0.19}_{-0.18}$   & 0.73$^{+0.12}_{-0.11}$ & 2.0  $\pm$ 0.0  & 1.47  &  D  &  114 \\ 
G113.29-29.69 & 6$^{+2}_{-2    }$         &0.7 $^{+0.1 }_{-0.3 }$ & 0.28$^{+0.07}_{-0.07}$  & 0.07 $^{+0.01  }_{-0.01  }$ & 1.1$^{+0.3}_{-0.3}$ & 0.90$^{+0.16}_{-0.16}$   & 0.69$^{+0.18}_{-0.17}$ & 1.0  $\pm$ 0.0  & -0.06 &  M  &   55 \\ 
G113.91-37.01 & 7$^{+2}_{-4    }$         &8   $^{+1   }_{-4.  }$ & 0.23$^{+0.03}_{-0.04}$  & 0.34 $^{+0.08  }_{-0.09  }$ & 0.6$^{+0.2}_{-0.2}$ & 0.40$^{+0.13}_{-0.12}$   & 0.86$^{+0.06}_{-0.05}$ & 1.7  $\pm$ 0.5  & 0.74  &  D  &   97 \\ 
G114.79-33.71 & 5$^{+2}_{-2    }$ 	  &0.7 $^{+0.3 }_{-0.6 }$ & 0.24$^{+0.06}_{-0.06}$  & 0.052$^{+0.001 }_{-0.011 }$ & 1.3$^{+0.3}_{-0.3}$ & 1.09$^{+0.18}_{-0.18}$   & 0.80$^{+0.10}_{-0.09}$ & 1.2  $\pm$ 0.4  & -0.07 &  M  &   54 \\ 
G124.20-36.48* & 120$^{+50}_{-50   }$      &160 $^{+60  }_{-60  }$ & 0.30$^{+0.05}_{-0.05}$  & 0.9  $^{+0.2   }_{-0.2   }$ & 1.3$^{+0.2}_{-0.2}$ & 0.67$^{+0.10}_{-0.09}$   & 0.55$^{+0.18}_{-0.17}$ & 2.0  $\pm$ 0.0  & 1.60  &  D  &  118 \\  
G143.26+65.24 & 14$^{+4}_{-6    }$        &18  $^{+6   }_{-9   }$ & 0.24$^{+0.03}_{-0.03}$  & 0.28 $^{+0.05  }_{-0.06  }$ & 1.0$^{+0.2}_{-0.2}$ & 0.74$^{+0.16}_{-0.15}$   & 0.77$^{+0.09}_{-0.08}$ & 1.7  $\pm$ 0.5  & 0.87  &  D  &  101 \\
G149.39-36.84 & 4$^{+2}_{-2    }$         &2.7 $^{+1.1 }_{-1.9 }$ & 0.20$^{+0.04}_{-0.04}$  & 0.13 $^{+0.01  }_{-0.03  }$ & 1.0$^{+0.3}_{-0.3}$ & 0.80$^{+0.16}_{-0.15}$   & 0.84$^{+0.08}_{-0.07}$ & 1.2  $\pm$ 0.4  & 0.37  &  M  &   76 \\
G155.27-68.42 & 4$^{+1}_{-2    }$    	  &3   $^{+4   }_{-1   }$ & 0.25$^{+0.05}_{-0.07}$  & 0.27 $^{+0.03  }_{-0.07  }$ & 0.5$^{+0.3}_{-0.3}$ & 0.43$^{+0.15}_{-0.13}$   & 0.84$^{+0.09}_{-0.06}$ & 1.2  $\pm$ 0.4  & 0.44  &  M  &   80 \\
G159.91-73.50 & 3.0$^{+0.1}_{-0.8  }$     &2.4 $^{+0.6 }_{-1.4 }$ & 0.31$^{+0.04}_{-0.04}$  & 0.19 $^{+0.03  }_{-0.04  }$ & 1.2$^{+0.3}_{-0.3}$ & 0.96$^{+0.13}_{-0.12}$   & 0.85$^{+0.06}_{-0.04}$ & 1.0  $\pm$ 0.0  & 0.17  &  M  &   65 \\ 
G172.74+65.30 & 16$^{+8}_{-9    }$        &0.3 $^{+0.3 }_{-0.1 }$ & 0.29$^{+0.08}_{-0.08}$  & 0.15 $^{+0.04  }_{-0.05  }$ & 1.3$^{+0.3}_{-0.3}$ & 0.99$^{+0.17}_{-0.17}$   & 0.67$^{+0.19}_{-0.19}$ & 1.7  $\pm$ 0.5  & 0.14  &  M  &   62 \\ 
G172.98-53.55 & 8$^{+2}_{-3    }$         &4   $^{+1   }_{-2   }$ & 0.23$^{+0.03}_{-0.04}$  & 0.09 $^{+0.02  }_{-0.04  }$ & 0.6$^{+0.2}_{-0.2}$ & 0.40$^{+0.12}_{-0.11}$   & 0.79$^{+0.10}_{-0.08}$ & 1.2  $\pm$ 0.4  & 0.36  &  M  &   74 \\ 
G179.09+60.12 & 0.6$^{+0.3}_{-0.4  }$     &0.3 $^{+0.6 }_{-0.3 }$ & 0.65$^{+0.06}_{-0.07}$  & 0.04 $^{+0.02  }_{-0.02  }$ & 0.8$^{+0.2}_{-0.1}$ & 0.67$^{+0.08}_{-0.08}$   & 0.81$^{+0.14}_{-0.12}$ & 0.7  $\pm$ 0.5  & -1.09 &  M  &   18 \\ 
G186.37+37.26 & 3.5$^{+0.4}_{-0.9  }$ 	  &0.7 $^{+0.3 }_{-0.6 }$ & 0.32$^{+0.03}_{-0.04}$  & 0.074$^{+0.001 }_{-0.013 }$ & 0.8$^{+0.2}_{-0.2}$ & 0.60$^{+0.12}_{-0.12}$   & 0.82$^{+0.07}_{-0.05}$ & 0.8  $\pm$ 0.4  & -0.20 &  M  &   49 \\ 
G187.53+21.92 & 1.4$^{+0.5}_{-0.8  }$ 	  &2.3 $^{+0.7 }_{-1.5 }$ & 0.48$^{+0.08}_{-0.09}$  & 0.031$^{+0.010 }_{-0.004 }$ & 0.9$^{+0.3}_{-0.3}$ & 0.77$^{+0.12}_{-0.11}$   & 0.83$^{+0.11}_{-0.09}$ & 0.3  $\pm$ 0.5  & -0.62 &  M  &   28 \\ 
G192.18+56.12 & 7$^{+3}_{-4    }$         &1.6 $^{+0.1 }_{-1.0 }$ & 0.26$^{+0.07}_{-0.08}$  & 0.10 $^{+0.03  }_{-0.05  }$ & 1.2$^{+0.3}_{-0.3}$ & 1.1 $^{+0.2 }_{-0.2 }$   & 0.76$^{+0.15}_{-0.14}$ & 1.0  $\pm$ 0.0  & 0.19  &  M  &   66 \\ 
G195.75-24.32 & 0.10$^{+0.01}_{-0.08 }$   &1.5 $^{+0.1 }_{-0.7 }$ & 0.20$^{+0.03}_{-0.04}$  & 0.11 $^{+0.02  }_{-0.04  }$ & 1.1$^{+0.3}_{-0.3}$ & 0.88$^{+0.13}_{-0.12}$   & 0.96$^{+0.03}_{-0.01}$ & 2.0  $\pm$ 0.0  & -0.28 &  M  &   47 \\ 
G201.50-27.31 & 3.2$^{+1.0}_{-1.9  }$     &0.1 $^{+0.4 }_{-0.1 }$ & 0.32$^{+0.03}_{-0.04}$  & 0.10 $^{+0.01  }_{-0.04  }$ & 0.4$^{+0.2}_{-0.2}$ & 0.27$^{+0.13}_{-0.11}$   & 0.82$^{+0.10}_{-0.07}$ & 1.0  $\pm$ 0.0  & -0.37 &  M  &   43 \\ 
G204.10+16.51 & 1.4$^{+0.9}_{-1.1  }$     &0.9 $^{+0.1 }_{-0.5 }$ & 0.34$^{+0.07}_{-0.07}$  & 0.15 $^{+0.03  }_{-0.04  }$ & 1.1$^{+0.3}_{-0.2}$ & 0.83$^{+0.14}_{-0.13}$   & 0.84$^{+0.10}_{-0.08}$ & 1.0  $\pm$ 0.0  & -0.17 &  M  &   51 \\ 
G205.93-39.46 & 4.8$^{+0.1}_{-1.3  }$     &0.17$^{+0.08}_{-0.14}$ & 0.38$^{+0.04}_{-0.05}$  & 0.06 $^{+0.01  }_{-0.02  }$ & 0.9$^{+0.2}_{-0.2}$ & 0.58$^{+0.13}_{-0.11}$   & 0.75$^{+0.11}_{-0.07}$ & 1.5  $\pm$ 0.5  & -0.47 &  M  &   37 \\ 
G206.45+13.89 & 17$^{+3}_{-6    }$        &13.1$^{+1.5 }_{-1.9 }$ & 0.32$^{+0.04}_{-0.05}$  & 0.28 $^{+0.04  }_{-0.07  }$ & 0.6$^{+0.3}_{-0.2}$ & 0.49$^{+0.13}_{-0.12}$   & 0.63$^{+0.17}_{-0.14}$ & 1.2  $\pm$ 0.4  & 0.70  &  M  &   90 \\ 
G207.88+81.31 & 19$^{+7}_{-10.  }$    	  &40  $^{+20  }_{-20  }$ & 0.24$^{+0.04}_{-0.05}$  & 0.12 $^{+0.03  }_{-0.05  }$ & 0.7$^{+0.3}_{-0.2}$ & 0.58$^{+0.14}_{-0.13}$   & 0.70$^{+0.15}_{-0.14}$ & 1.2  $\pm$ 0.4  & 0.81  &  M  &   91 \\
G208.80-30.67 & 24$^{+6}_{-8	}$    	  &20  $^{+10  }_{-11  }$ & 0.15$^{+0.02}_{-0.03}$  & 0.7  $^{+0.2   }_{-0.2   }$ & 1.0$^{+0.2}_{-0.2}$ & 0.55$^{+0.12}_{-0.12}$   & 0.84$^{+0.03}_{-0.02}$ & 1.8  $\pm$ 0.4  & 1.43  &  D  &  113 \\
G210.64+17.09 & 5$^{+1}_{-3    }$         &10  $^{+2.  }_{-7   }$ & 0.19$^{+0.04}_{-0.05}$  & 0.19 $^{+0.03  }_{-0.07  }$ & 0.5$^{+0.4}_{-0.4}$ & 0.44$^{+0.17}_{-0.16}$   & 0.88$^{+0.09}_{-0.04}$ & 1.3  $\pm$ 0.5  & 0.68  &  M  &   88 \\ 
G216.62+47.00 & 4$^{+1}_{-3    }$ 	  &16  $^{+1.  }_{-12. }$ & 0.40$^{+0.08}_{-0.01}$  & 0.24 $^{+0.09  }_{-0.16  }$ & 1.1$^{+0.4}_{-0.4}$ & 1.04$^{+0.17}_{-0.15}$   & 0.77$^{+0.17}_{-0.10}$ & 1.2  $\pm$ 0.4  & 0.35  &  M  &   73 \\ 
G217.09+40.15 & 1.8$^{+0.8}_{-1.0  }$     &2.4 $^{+0.3 }_{-1.0 }$ & 0.42$^{+0.06}_{-0.07}$  & 0.17 $^{+0.02  }_{-0.03  }$ & 1.0$^{+0.2}_{-0.2}$ & 0.75$^{+0.12}_{-0.11}$   & 0.85$^{+0.09}_{-0.07}$ & 1.0  $\pm$ 0.0  & -0.10 &  M  &   53 \\
G217.40+10.88 & 0.4$^{+0.1}_{-0.2  }$     &0.09$^{+0.05}_{-0.06}$ & 0.57$^{+0.09}_{-0.09}$  & 0.018$^{+0.004 }_{-0.001 }$ & 0.8$^{+0.2}_{-0.2}$ & 0.62$^{+0.11}_{-0.10}$   & 0.89$^{+0.09}_{-0.06}$ & 0.3  $\pm$ 0.5  & -1.41 &  R  &    8 \\
G218.59+71.31 & 28$^{+14}_{-16.  }$       &1.4 $^{+0.5 }_{-1.2 }$ & 0.10$^{+0.02}_{-0.02}$  & 0.10 $^{+0.02  }_{-0.01  }$ & 1.5$^{+0.2}_{-0.2}$ & 1.2 $^{+0.2 }_{-0.2 }$   & 0.73$^{+0.15}_{-0.13}$ & 1.8  $\pm$ 0.4  & 0.89  &  D  &  102 \\
G218.81+35.51 & 40$^{+20}_{-20   }$  	  & 50 $^{+20. }_{-30. }$ & 0.40$^{+0.09}_{-0.01}$  & 0.42 $^{+0.15  }_{-0.17  }$ & 1.2$^{+0.3}_{-0.3}$ & 0.95$^{+0.13}_{-0.12}$   & 0.6 $^{+0.2 }_{-0.2 }$ & 1.5  $\pm$ 0.5  & 0.96  &  D  &  104 \\ 
G224.00+69.33 & 11$^{+4}_{-5	 }$  	  &3.6 $^{+0.7 }_{-1.8 }$ & 0.29$^{+0.04}_{-0.04}$  & 0.05 $^{+0.01  }_{-0.02  }$ & 0.9$^{+0.2}_{-0.2}$ & 0.70$^{+0.13}_{-0.12}$   & 0.76$^{+0.12}_{-0.10}$ & 1.0  $\pm$ 0.0  & 0.13  &  M  &   61 \\ 
G225.93-19.99 & 63$^{+11}_{-18.  }$  	  &60  $^{+10  }_{-20  }$ & 0.24$^{+0.02}_{-0.03}$  & 0.85 $^{+0.07  }_{-0.10  }$ & 1.1$^{+0.2}_{-0.2}$ & 0.71$^{+0.17}_{-0.15}$   & 0.75$^{+0.09}_{-0.04}$ & 2.0  $\pm$ 0.0  & 1.49  &  D  &  115 \\ 
G226.18+76.79 & 2.8$^{+0.3}_{-0.6  }$     &1.4 $^{+0.2 }_{-0.5 }$ & 0.44$^{+0.04}_{-0.04}$  & 0.04 $^{+0.01  }_{-0.02  }$ & 0.8$^{+0.2}_{-0.2}$ & 0.59$^{+0.09}_{-0.09}$   & 0.78$^{+0.08}_{-0.07}$ & 0.7  $\pm$ 0.5  & -0.47 &  M  &   36 \\ 
G228.16+75.20 & 3.3$^{+0.5}_{-0.7  }$     &2.5 $^{+0.4 }_{-1.9 }$ & 0.21$^{+0.03}_{-0.03}$  & 0.30 $^{+0.06  }_{-0.09  }$ & 0.3$^{+0.2}_{-0.2}$ & 0.22$^{+0.13}_{-0.11}$   & 0.93$^{+0.01}_{-0.01}$ & 1.2  $\pm$ 0.4  & 0.50  &  M  &   83 \\ 
G229.74+77.96 & 27 $^{+9}_{-12.  }$       &16  $^{+4   }_{-7   }$ & 0.21$^{+0.03}_{-0.04}$  & 0.35 $^{+0.04  }_{-0.07  }$ & 1.1$^{+0.3}_{-0.3}$ & 0.86$^{+0.16}_{-0.15}$   & 0.67$^{+0.15}_{-0.13}$ & 1.8  $\pm$ 0.4  & 1.07  &  D  &  109 \\ 
G238.69+63.26 & 3.0$^{+0.9}_{-1.3  }$     &4   $^{+2   }_{-2   }$ & 0.28$^{+0.04}_{-0.04}$  & 0.24 $^{+0.04  }_{-0.05  }$ & 0.9$^{+0.2}_{-0.2}$ & 0.62$^{+0.12}_{-0.11}$   & 0.88$^{+0.05}_{-0.04}$ & 1.3  $\pm$ 0.5  & 0.34  &  M  &   72 \\ 
G239.27-26.01 & 2.6$^{+0.7}_{-1.3  }$ 	  &1.5 $^{+0.1 }_{-0.9 }$ & 0.24$^{+0.02}_{-0.03}$  & 0.30 $^{+0.02  }_{-0.03  }$ & 0.6$^{+0.2}_{-0.2}$ & 0.40$^{+0.13}_{-0.12}$   & 0.89$^{+0.06}_{-0.04}$ & 1.5  $\pm$ 0.5  & 0.34  &  M  &   70 \\ 
G241.11-28.68 & 0.1$^{+0.7}_{-0.1  }$ 	  &3.2 $^{+3.3 }_{-0.3 }$ & 0.28$^{+0.04}_{-0.06}$  & 0.12 $^{+0.01  }_{-0.04  }$ & 0.8$^{+0.3}_{-0.3}$ & 0.66$^{+0.17}_{-0.15}$   & 0.93$^{+0.04}_{-0.01}$ & 1.0  $\pm$ 0.0  & -0.35 &  M  &   44 \\ 
G243.15-73.84 & 2.6$^{+0.7}_{-1.3  }$ 	  &0.7 $^{+0.1 }_{-0.6 }$ & 0.16$^{+0.01}_{-0.02}$  & 0.22 $^{+0.02  }_{-0.04  }$ & 0.4$^{+0.2}_{-0.2}$ & 0.32$^{+0.14}_{-0.13}$   & 0.90$^{+0.05}_{-0.03}$ & 1.5  $\pm$ 0.5  & 0.39  &  M  &   77 \\ 
G243.64+67.74 & 1.4$^{+0.2}_{-0.4  }$ 	  &0.23$^{+0.06}_{-0.18}$ & 0.27$^{+0.04}_{-0.05}$  & 0.38 $^{+0.07  }_{-0.08  }$ & 1.3$^{+0.2}_{-0.2}$ & 0.94$^{+0.16}_{-0.15}$   & 0.95$^{+0.01}_{-0.01}$ & 1.5  $\pm$ 0.5  & 0.01  &  M  &   57 \\ 
G259.98-63.43 & 0.4$^{+0.2}_{-0.1  }$ 	  &0.5 $^{+0.3 }_{-0.5 }$ & 0.44$^{+0.04}_{-0.05}$  & 0.06 $^{+0.01  }_{-0.02  }$ & 1.0$^{+0.2}_{-0.2}$ & 0.75$^{+0.12}_{-0.11}$   & 0.89$^{+0.06}_{-0.02}$ & 1.0  $\pm$ 0.0  & -0.78 &  M  &   23 \\ 
G262.27-35.38 & 15 $^{+5}_{-6    }$       &2.2 $^{+0.5 }_{-0.8 }$ & 0.14$^{+0.02}_{-0.03}$  & 0.35 $^{+0.08  }_{-0.10  }$ & 1.3$^{+0.2}_{-0.2}$ & 1.01$^{+0.13}_{-0.12}$   & 0.82$^{+0.07}_{-0.06}$ & 1.8  $\pm$ 0.4  & 0.97  &  D  &  105 \\ 
G262.73-40.92 & 2.6$^{+0.7}_{-1.4  }$ 	  &1.1 $^{+0.6 }_{-0.5 }$ & 0.38$^{+0.04}_{-0.05}$  & 0.043$^{+0.001 }_{-0.013 }$ & 0.6$^{+0.3}_{-0.2}$ & 0.49$^{+0.14}_{-0.12}$   & 0.80$^{+0.12}_{-0.09}$ & 0.8  $\pm$ 0.4  & -0.41 &  M  &   40 \\ 
G263.68-22.55 & 2.4$^{+0.2}_{-0.3  }$ 	  &0.1 $^{+0.3 }_{-0.1 }$ & 0.41$^{+0.06}_{-0.06}$  & 0.10 $^{+0.01  }_{-0.02  }$ & 1.2$^{+0.3}_{-0.3}$ & 0.97$^{+0.13}_{-0.12}$   & 0.78$^{+0.11}_{-0.09}$ & 1.0  $\pm$ 0.0  & -0.55 &  M  &   29 \\ 
G266.04-21.25 & 1.6$^{+0.1}_{-0.4  }$ 	  &1.7 $^{+0.3 }_{-0.8 }$ & 0.27$^{+0.02}_{-0.03}$  & 0.13 $^{+0.02  }_{-0.03  }$ & 0.8$^{+0.2}_{-0.2}$ & 0.59$^{+0.11}_{-0.10}$   & 0.85$^{+0.06}_{-0.04}$ & 2.0  $\pm$ 0.0  & 0.02  &  M  &   58 \\ 
G266.83+25.08 & 1.0$^{+0.1}_{-0.2  }$ 	  &0.03$^{+0.02}_{-0.03}$ & 0.55$^{+0.04}_{-0.04}$  & 0.057$^{+0.003 }_{-0.010 }$ & 0.7$^{+0.1}_{-0.1}$ & 0.47$^{+0.08}_{-0.07}$   & 0.83$^{+0.06}_{-0.04}$ & 0.2  $\pm$ 0.4  & -1.12 &  R  &   12 \\ 
G271.18-30.95 & 0.06$^{+0.01}_{-0.04 }$   &0.38$^{+0.18}_{-0.07}$ & 0.55$^{+0.04}_{-0.04}$  & 0.016$^{+0.001 }_{-0.016 }$ & 0.3$^{+0.1}_{-0.1}$ & 0.21$^{+0.07}_{-0.07}$   & 0.96$^{+0.02}_{-0.01}$ & 0.0  $\pm$ 0.0  & -1.53 &  R  &    6 \\ 
G273.59+63.27 & 17$^{+12}_{-13.  }$       &0.2 $^{+0.3 }_{-0.2 }$ & 0.19$^{+0.08}_{-0.09}$  & 0.14 $^{+0.08  }_{-0.09  }$ & 1.4$^{+0.5}_{-0.5}$ & 1.3 $^{+0.2 }_{-0.2 }$   & 0.66$^{+0.26}_{-0.24}$ & 1.7  $\pm$ 0.5  & 0.32  &  M  &   69 \\ 
G277.76-51.74 & 5$^{+1}_{-2    }$         &15  $^{+1   }_{-5   }$ & 0.14$^{+0.02}_{-0.02}$  & 0.37 $^{+0.09  }_{-0.11  }$ & 0.6$^{+0.2}_{-0.2}$ & 0.40$^{+0.13}_{-0.12}$   & 0.85$^{+0.08}_{-0.06}$ & 1.7  $\pm$ 0.5  & 1.05  &  D  &  107 \\ 
G278.58+39.16 & 12$^{+4}_{-5    }$        &12  $^{+3   }_{-6   }$ & 0.32$^{+0.05}_{-0.06}$  & 0.37 $^{+0.09  }_{-0.11  }$ & 1.2$^{+0.3}_{-0.3}$ & 0.89$^{+0.13}_{-0.11}$   & 0.74$^{+0.13}_{-0.10}$ & 1.3  $\pm$ 0.5  & 0.70  &  M  &   89 \\ 
G283.91+73.87* & 7$^{+7}_{-7    }$         &30  $^{+30  }_{-30  }$ & 0.80$^{+0.05}_{-0.05}$  &  1.0  $^{+0.9   }_{-0.9  }$ & 1.3$^{+0.4 }_{0.4}$ & 1.01$^{+0.10 }_{0.10}$   & 0.84 $^{+0.07}_{-0.07}$& 2.0  $\pm$ 0.0  & 0.47  &  M  &   81  \\  
G284.41+52.45 & 0.35$^{+0.06}_{-0.05 }$   &0.23$^{+0.11}_{-0.19}$ & 0.40$^{+0.02}_{-0.02}$  & 0.017$^{+0.002 }_{-0.001 }$ & 0.4$^{+0.1}_{-0.1}$ & 0.20$^{+0.06}_{-0.06}$   & 0.93$^{+0.02}_{-0.01}$ & 0.7  $\pm$ 0.5  & -1.14 &  M  &   17 \\ 
G285.63+72.75 & 6$^{+3}_{-4    }$         &10  $^{+6   }_{-7   }$ & 0.30$^{+0.07}_{-0.08}$  & 0.07 $^{+0.003 }_{-0.01  }$ & 0.9$^{+0.3}_{-0.3}$ & 0.68$^{+0.11}_{-0.11}$   & 0.74$^{+0.17}_{-0.16}$ & 1.0  $\pm$ 0.0  & 0.22  &  M  &   68 \\ 
G286.98+32.90 & 5$^{+1}_{-3    }$         &6   $^{+1   }_{-4   }$ & 0.22$^{+0.04}_{-0.04}$  & 0.17 $^{+0.010 }_{-0.04  }$ & 1.1$^{+0.3}_{-0.3}$ & 0.99$^{+0.15}_{-0.14}$   & 0.90$^{+0.05}_{-0.01}$ & 1.3  $\pm$ 0.5  & 0.51  &  M  &   84 \\ 
G287.46+81.12 & 2.6$^{+0.3}_{-0.9  }$     &1.1 $^{+0.3 }_{-0.8 }$ & 0.17$^{+0.04}_{-0.05}$  & 0.16 $^{+0.02  }_{-0.04  }$ & 1.5$^{+0.3}_{-0.3}$ & 1.33$^{+0.19}_{-0.19}$   & 0.87$^{+0.08}_{-0.06}$ & 1.3  $\pm$ 0.5  & 0.34  &  M  &   71 \\	
G313.33+61.13 & 0.27$^{+0.02}_{-0.05 }$	  &0.04$^{+0.11}_{-0.05}$ & 0.55$^{+0.04}_{-0.04}$  & 0.042$^{+0.001 }_{-0.004 }$ & 0.6$^{+0.1}_{-0.1}$ & 0.45$^{+0.06}_{-0.06}$   & 0.88$^{+0.05}_{-0.04}$ & 0.0  $\pm$ 0.0  & -1.35 &  R  &    9 \\ 
G313.88-17.11 & 0.3$^{+0.2}_{-0.3  }$ 	  &0.3 $^{+0.3 }_{-0.1 }$ & 0.50$^{+0.07}_{-0.07}$  & 0.034$^{+0.010 }_{-0.002 }$ & 0.9$^{+0.2}_{-0.2}$ & 0.72$^{+0.11}_{-0.10}$   & 0.86$^{+0.12}_{-0.10}$ & 0.2  $\pm$ 0.4  & -1.09 &  R  &   14 \\	
G324.04+48.79 & 0.37$^{+0.17}_{-0.03 }$	  &0.13$^{+0.02}_{-0.08}$ & 0.60$^{+0.03}_{-0.03}$  & 0.023$^{+0.010 }_{-0.002 }$ & 0.5$^{+0.1}_{-0.1}$ & 0.30$^{+0.06}_{-0.06}$   & 0.88$^{+0.03}_{-0.01}$ & 0.2  $\pm$ 0.4  & -1.35 &  R  &   10 \\ 
G325.70+17.34 & 1.1$^{+0.8}_{-1.0  }$ 	  &0.03$^{+0.05}_{-0.04}$ & 0.28$^{+0.05}_{-0.06}$  & 0.17 $^{+0.03  }_{-0.04  }$ & 0.6$^{+0.2}_{-0.2}$ & 0.44$^{+0.08}_{-0.07}$   & 0.88$^{+0.09}_{-0.08}$ & 1.2  $\pm$ 0.4  & -0.48 &  M  &   34  \\  
G339.63-69.34 & 0.13$^{+0.17}_{-0.02 }$   &0.14$^{+0.02}_{-0.13}$ & 0.65$^{+0.04}_{-0.05}$  & 0.007$^{+0.010 }_{-0.001 }$ & 0.4$^{+0.2}_{-0.1}$ & 0.30$^{+0.08}_{-0.07}$   & 0.93$^{+0.03}_{-0.01}$ & 0.0  $\pm$ 0.0  & -1.82 &  R  &    2  \\
G340.36+60.58 & 0.07$^{+0.18}_{-0.09 }$   &0.01$^{+0.03}_{-0.02}$ & 0.62$^{+0.04}_{-0.05}$  & 0.017$^{+0.012 }_{-0.010 }$ & 0.6$^{+0.1}_{-0.1}$ & 0.45$^{+0.05}_{-0.05}$   & 0.93$^{+0.01}_{-0.00}$ & 0.0  $\pm$ 0.0  & -1.99 &  R  &    1  \\
G340.94+35.07 & 1.3$^{+0.5}_{-0.7}$       &0.2 $^{+0.6 }_{-0.2 }$ & 0.57$^{+0.01}_{-0.01}$  & 0.07 $^{+0.02  }_{-0.03  }$ & 0.7$^{+0.2}_{-0.2}$ & 0.56$^{+0.05}_{-0.05}$   & 0.80$^{+0.16}_{-0.14}$ & 0.7  $\pm$ 0.5  & -0.82 &  M  &   20 \\ 
G346.61+35.06 & 23$^{+12}_{-16}$          &40  $^{+20  }_{-30. }$ & 0.14$^{+0.04}_{-0.05}$  & 0.6  $^{+0.3   }_{-0.3   }$ & 1.4$^{+0.5}_{-0.5}$ & 1.26$^{+0.20}_{-0.19}$   & 0.79$^{+0.12}_{-0.10}$ & 1.8  $\pm$ 0.4  & 1.51  &  D  &  116 \\ 
G349.46-59.95 & 1.3$^{+0.3}_{-0.5}$       &0.3 $^{+0.3 }_{-0.1 }$ & 0.44$^{+0.03}_{-0.03}$  & 0.051$^{+0.010 }_{-0.014 }$ & 0.6$^{+0.2}_{-0.2}$ & 0.42$^{+0.09}_{-0.08}$   & 0.86$^{+0.06}_{-0.05}$ & 0.8  $\pm$ 0.4  & -0.71 &  M  &   25 \\ 
\hline\hline

\end{longtable}

\end{ThreePartTable}

\end{landscape}

\twocolumn


\section{Other morphological parameters}\label{Other_parameters}
\FloatBarrier
\begin{figure}
    \hspace{-0.8cm}
    \includegraphics[scale=0.6]{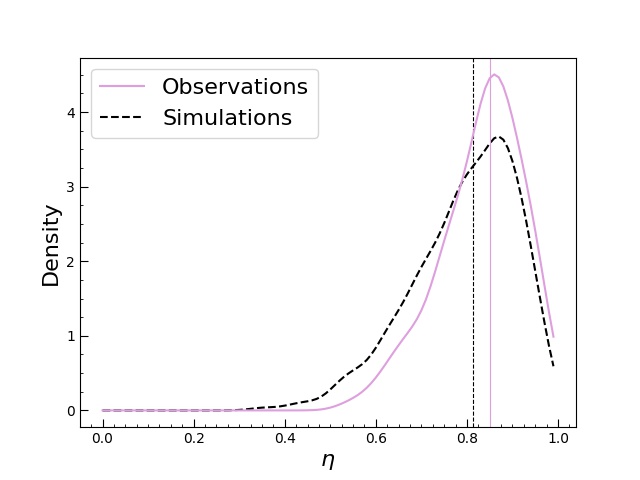}
    \caption{Distribution of $\eta$ for the observed (violet solid line) and simulated (black dashed line) samples. The vertical lines represent the medians of the two distributions.}
    \label{fig:elll}
\end{figure}

\subsection{Ellipticity}
The ellipticity parameter is able to describe the shape of the surface brightness distribution and could provide a link to the dynamical state. Indeed, relaxed systems are expected to be rounder than the disturbed clusters. The ellipticity parameter is defined as:
\begin{equation}
    \eta = 1-\frac{b}{a}
\end{equation}
where, $a$ and $b$ are respectively the major and the minor semi-axis. Here, for simplicity, we consider only the axial ratio (i.e. $b/a$), and consequently $\eta$ takes values from 0 (elliptical shape) to 1 (circualr shape). \\
Observing the correlation of $\eta$ with the other parameters, we noticed a strong correlation between $\eta$ -- $P_{20}$ ($r$ = -0.81), which stresses out that these two parameters are representing the same quantity, i.e. the cluster ellipticity. For this reason, we decide to consider only $P_{20}$ for the rest of the analysis and to report in this Appendix the correlations of $\eta$ with the other parameters. For the sake of completeness, in Fig.~\ref{fig:elll} is reported the distribution of $\eta$ for the observed and simulated samples.

\subsection{Asymmetry and Smoothness } 
In this Subsection we present the analysis realised using the Asymmetry and the Smoothness, which are two parameters originally used to study the morphology of galaxies. 
The asymmetry parameter, $A$, is a measure of how the light distribution differs from a spherically symmetric distribution and is computed by subtracting from the original image an image rotate by \ang{180} \citep{conselice2000, Lotz2004}:
\begin{equation}
    A=\frac{\sum_{i,j}\left|\text{I}(i,j)-\text{I}_{180}(i,j) \right|}{\sum_{i,j}\left|\text{I}(i,j) \right|}
\end{equation}
where I is the cluster’s image, I$_{180}$ is the image rotated by \ang{180} around the pixel corresponding to the X-ray peak. 

The smoothness parameter, $S$, is obtained by subtracting a Gaussian smoothed image, $I_s$ from the original one, $I$ \citep[see also,][]{Lotz2004, conselice2003}:
\begin{equation}
    S=\frac{\sum_{i,j}\left|\text{I}(i,j)-\text{I}_{\text{s}}(i,j)\right|}{\sum_{i,j}\left|\text{I}(i,j) \right|}
\end{equation}
\begin{equation}
    S_{\text{bkg}}= \frac{\sum_{i,j}\left|\text{B}(i,j)-\text{B}_{\text{s}}(i,j) \right|}{\sum_{i,j}\left|\text{I}(i,j) \right|}
\end{equation}
In order to compare $S$ values of clusters spanning a wide range of redshifts, we decided to suppress inhomogeneities smaller than the minimal scale resolved in high-redshift images, i.e. $\sim$\,100 kpc. Thanks to this condition, $S$ is sensible to the same sub-clusters scales, independently from the redshift of the considered objects.     

\begin{figure*}
    \centering
    \includegraphics[scale=0.6]{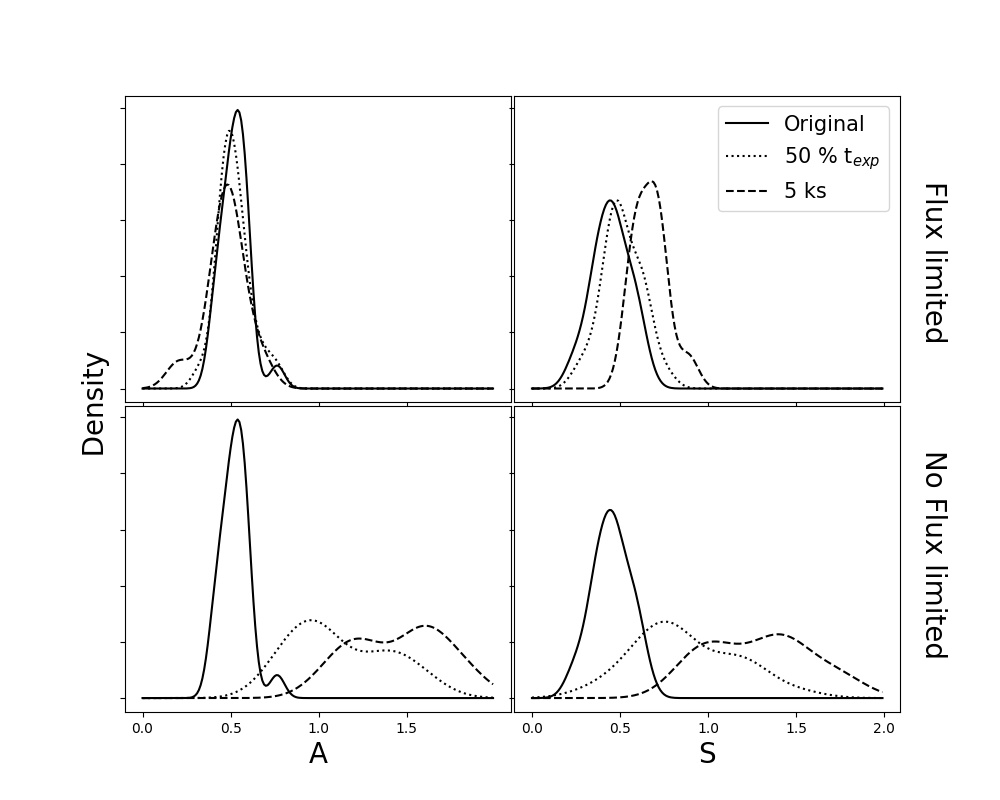}
    \caption{Comparison between the distribution of $A$ and $S$ computed using images with the original exposure time (solid line), with halved exposure time (dotted line) or with an exposure time of 5 ks. The bottom plots show the results obtained when no threshold is applied, the upper plots instead show the results obtained when only pixels with value higher than 0 are considered. 
    }
    \label{fig:A_S_50p}
\end{figure*}

We then computed this two parameters for the CHEX-MATE and simulated sample and we report the results in Table \ref{tab:AS}. In particular, for what concerns the simulated sample, the values reported are obtained with the same method adopted in Sect. \ref{morpho_sim} (i.e., building 10$^4$ randomly extracted subsamples, computing the median value of the Spearman coefficient and the corresponding standard deviation). It is possible to observe that the observed median and Spearman coefficient differ from the simulated one. Furthermore, a very strong correlation is observed between $A$ -- $S$, with $r$ = 0.94. The discrepancies between the observed and simulated distributions of the asymmetry and smoothness could be an indications that these two parameters are strongly influenced by the signal to noise ratio (S/N) which is equal to S/N $\sim$ 150 for the observations and S/N $\sim$ $\infty$ for the simulations. To test this hypothesis, we considered a sub-sample of 20 CHEX-MATE clusters and we repeat the morphological analysis on images with both halved t$_{\text{exp}}$ and t$_{\text{exp}}=$ 5 ks. In Figure~\ref{fig:A_S_50p}, we show the comparison between the distribution of $A$ and $S$ computed using the original images (solid line) and the distributions obtained using images with a reduced exposure time (dotted line for $\sim$ 50 \% t$_{\text{ext}}$ and in dashed line for t $\sim$ 5 ks). As it is possible to notice, these two parameters show a shift in their distributions which increases as the exposure time decreases. 

Before drawing any conclusion, we make the following consideration: originally the Asymmetry and the Smoothness where defined to classify the morphology of galaxies in the optical band, where images are characterised by a larger number of photon counts. X-ray images instead, are characterised by a sparse emission, which means that close to pixel with counts also pixels without counts are present. When computing $A$ (or $S$), for each pixel, the absolute value of the difference between the original and the rotated (or smoothed) image is realised. The local difference between pixels with and without counts could thus result in high values of the Asymmetry (or Smoothness), even if the overall shape of the emission is still symmetric \citep[as already observed by][]{nurgaliev2017}. The difference observed is not due to a real decrease of the flux, but due to the sparse distribution of the emission. This consideration could explain why the Spearman coefficient of the couple $A$ and $S$ is so high: they are not quantifying the shape of the X-ray emission, but the number of pixels with no emission. We thus decide to recompute this two parameters by considering only the pixels with values higher than 0 and we report the new results in Table \ref{tab:AS}. The Spearman coefficient obtained for the couple $A$ -- $S$ is consistent with what obtained in \cite{Parekh2015}; however discrepancies are still present between the observed and simulated $S$. Given these result we prefer to not include this parameters in our analysis.

\begin{table*}[]
\centering
\caption{Results obtained for the Asymmetry and the Smoothness.} \label{tab:AS}
\begin{tabular}{@{}ccccccc@{}}
\toprule
          & median & c$_\text{{obs}}$ & w$_\text{{obs}}$ & P$_\text{{20, obs}}$ & P$_\text{{30, obs}}$ & S$_\text{{obs}}$ \\ \midrule
A$_\text{{obs}}$   &  0.93      & -0.64   & 0.63  &     0.54     & 0.53   &  0.94     \\
S$_\text{{obs}}$  &  0.70       & -0.53   & 0.46    &   0.44       &  0.45  &  --    \\ \midrule
          & median & c$_\text{{sim}}$ & w$_\text{{sim}}$ & P$_\text{{20, sim}}$ & P$_\text{{30, sim}}$ & S$_\text{{sim}}$ \\ \midrule
          A$_\text{{sim}}$ & 0.58 $\pm$ 0.03      & -0.57$\pm$ 0.06  & 0.86 $\pm$ 0.03   &  0.62 $\pm$ 0.06        & 0.60 $\pm$ 0.06    &  0.06 $\pm$ 0.01   \\
S$_\text{{sim}}$ &  0.065 $\pm$ 0.004   & 0.46 $\pm$ 0.08  &  -0.09 $\pm$ 0.09 & -0.06 $\pm$ 0.09         &   -0.06 $\pm$ 0.09   &  --  \\ \midrule
& median & c$_\text{{obs}}$ & w$_\text{{obs}}$ & P$_\text{{20, obs}}$ & P$_\text{{30, obs}}$ & S$_\text{{obs}}$ \\ \midrule
 A$_\text{{flim}}$ &  0.87  & -0.47  & 0.49  &  0.38        & 0.44    &  0.54  \\
S$_\text{{flim}}$ & 1.11   & -0.60  &  0.52 & 0.40        &   0.44   &  --  \\ \bottomrule
\end{tabular}
\tablefoot{Second column: median of the observed and simulated Asymmetry and Smoothnes distributions. From the third to the seventh column: Spearman coefficient of the observed/simulated Asymmetry and smoothness with the other observed/simulated parameters.}
\end{table*}

\section{Morphological parameters at 0.5 $R_{500}$}\label{05analysis}
\begin{figure*}[h!]
    \centering
    \includegraphics[scale=0.45]{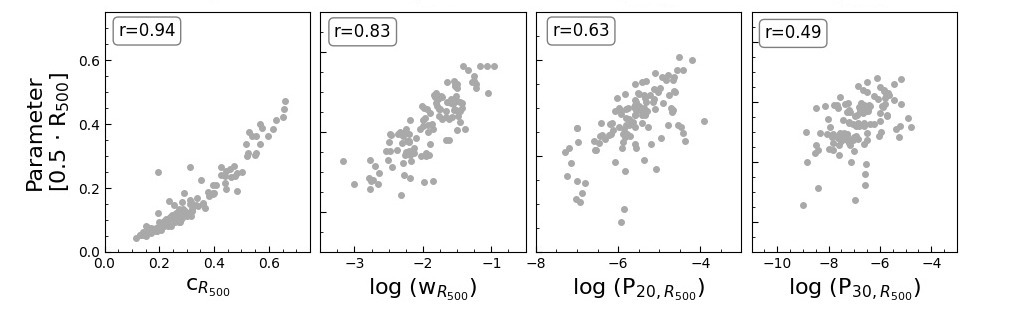}
    \caption{Correlations between the morphological parameters estimated inside $R_{500}$ (x-axis) and 0.5 $\cdot$ $R_{500}$ (y-axis).}
    \label{fig:05R}
\end{figure*}

\begin{figure*}[h!]
    \centering
    \includegraphics[scale=0.45]{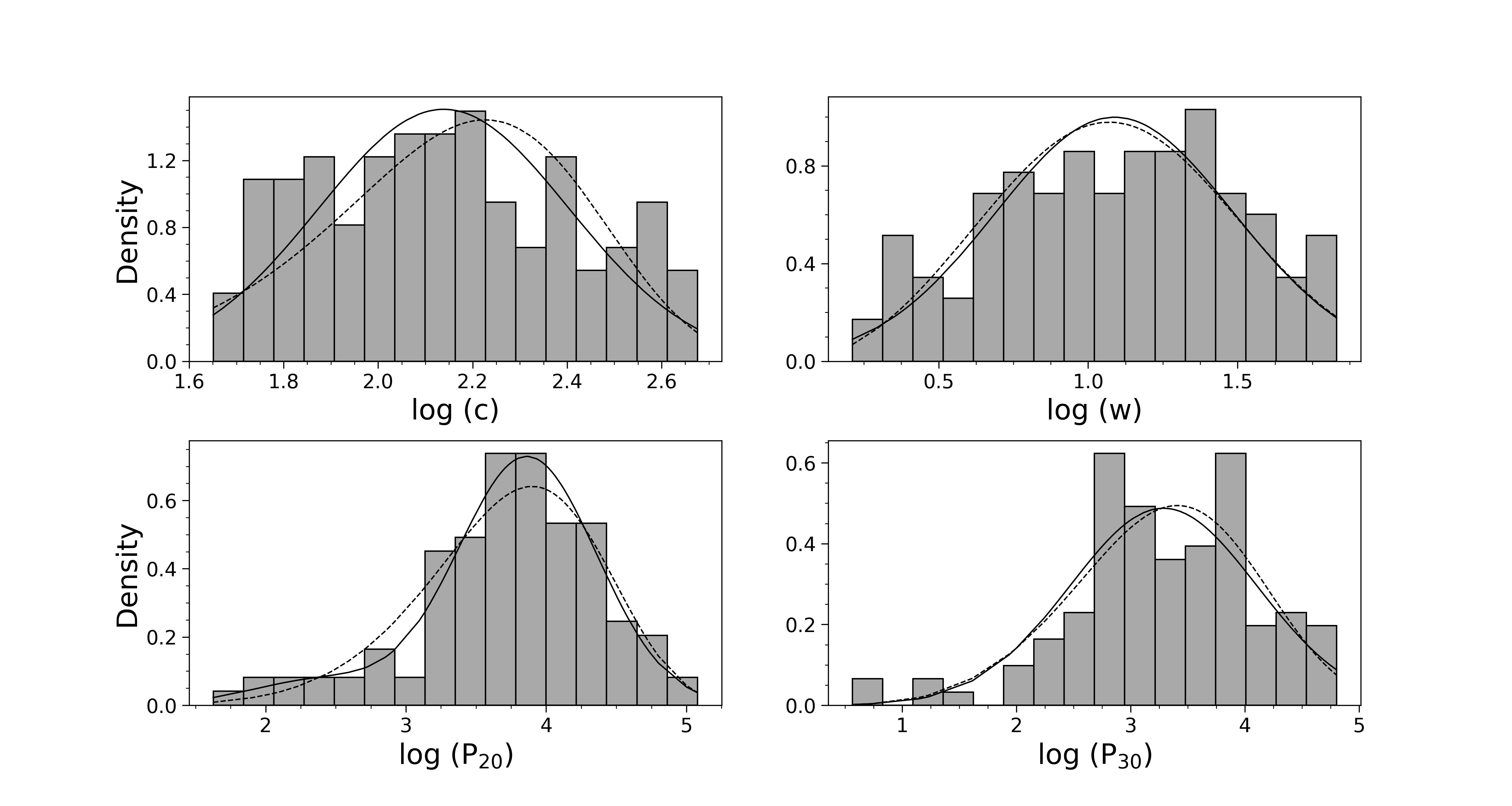}
    \caption{Distribution of $c$, $w$, $P_{20}$ and $P_{30}$ computed inside 0.5 $\cdot R_{500}$. The solid line represents the model obtained from the clustering analysis, while the dashed line represents the Weibull function. The parameters $c$, $w$, $P_{20}$ and $P_{30}$ were multiplied by 10$^3$, 10$^4$, 10$^9$ and 10$^{10}$, respectively.}
    \label{fig:quattro}
\end{figure*}

In this Section we report the result of the morphological analysis realised using a region of radius 0.5 $R_{500}$. First of all, we compared the values of the morphological parameters obtained using these new region, with the values estimated inside $R_{500}$. The results are shown in Fig. \ref{fig:05R}. It is possible to observe that strong correlations ($r$>0.5) are obtained for all the parameters. We then realised a cluster analysis as the one realised in Subsect. \ref{Sect:distributions}, and we present the results in Table \ref{BIC_val05} and in Fig. \ref{fig:quattro}. The Gaussian mixture model reveals the presence of a single component for all the parameters, with the exception of the $P_{20}$. For this latter indicator two components are identified, suggesting that a bimodality may characterise its distribution. However, it is possible to observe that the Weibull function has a highest BIC value. In particular, since the discrepancy between the Weibull and Gaussian BIC values is $\sim$ 6, it is possible to asses that the Weibull model is favoured (see Sec.\ref{Sect:distributions} for more details). Therefore, we concluded that the four parameters computed within 0.5 $R_{500}$ do not show sign of bimodality. Given the BIC values, we found that the single Gaussian component model is favoured in describing the distribution of the concentration. For the other parameter, it is instead not possible to unambiguously identify the best fit model.

\begin{table}
\centering
\renewcommand{\arraystretch}{1.3}
\caption{BIC values of the fit realised using the Weibull function, and the Gaussian mixture models.} \label{BIC_val05}
\begin{tabular}{lcc}
\hline
    & Weibull & \begin{tabular}[c]{@{}c@{}}Gaussian\\ mixture\tablefootnote{For $c$, $w$, $P_{30}$ is meant a single Gaussian component model, while for $P_{20}$ a double Gaussian component model.} \end{tabular} \\ \hline
c      & -37     & -30                                                      \\
w      & -123    & -125                                                       \\
P20    & -226    & -232                                                       \\
P30    & -288    & -289  \\ \hline                                                    
\end{tabular}

\tablefoot{The four morphological parameters were estimated within 0.5 $R_{500}$.}

\end{table}

\FloatBarrier


\section{Principal component analysis}\label{PCA_appendix}
\begin{figure*}[ht!]
    \hspace{1cm}
    \includegraphics[scale=0.46]{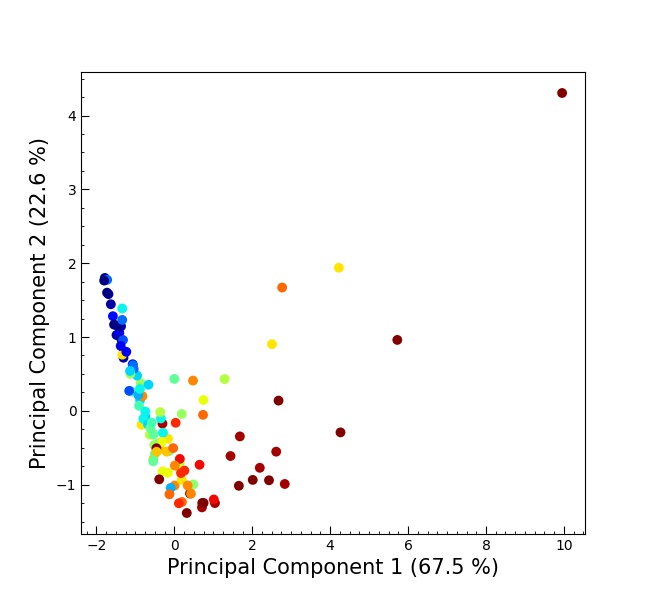}
    \includegraphics[scale=0.46]{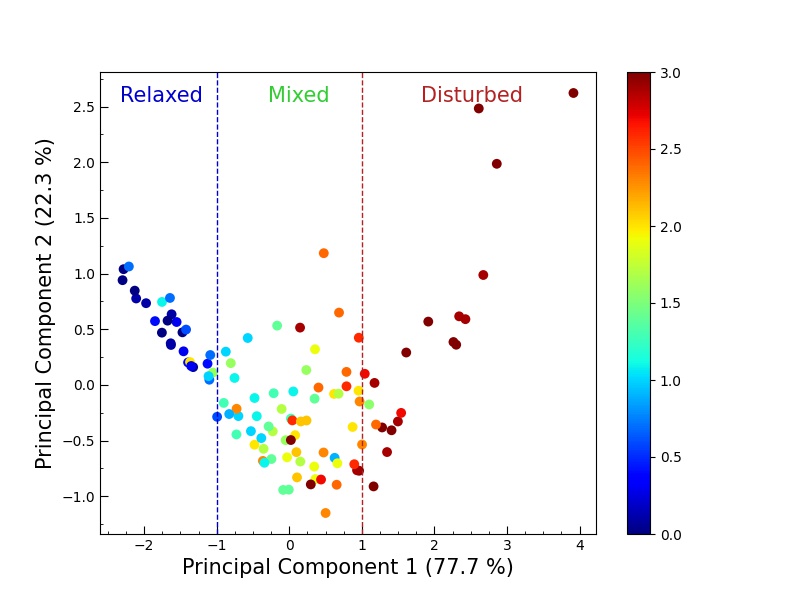}
    \caption{Results obtained by applying the PCA to the observations and using as input parameters $c$, $w$,$P_{20}$ and $P_{30}$ (left panel) or only $c$, $w$ (right panel). The colour scale represents the grade of relaxation of a system (with 0 being the most relaxed systems and 3 the most disturbed ones) defined by the visual classification (see Sect. \ref{visual classification}). }
    \label{fig:pca_observations}
\end{figure*}

Some parameters may show good correlations because they essentially provide the same type of information. In order to recognise redundancies and determine the minimal number of dimensions able to describe the properties of the observed clusters, we applied to our dataset the so called Principal Component Analysis, PCA \citep{Pearson1901,Hotelling1933, Jolliffe2011}. This procedure is used to represent an original set of $n_0$ mutually correlated random variables (in this case our four morphological parameters) with a smaller set ($n<n_0$) of independent hypothetical variables and allows to reduce the number of dimensions able to describe a dataset, without much loss of information. 

Before starting this analysis, we realised a standardisation of the dataset, which consists in re-scaling the values of each parameter in order to obtain a standard normal distribution, with a mean of zero and a standard deviation of one. We then computed the eigenvectors and eigenvalues of the covariance matrix; the eigenvectors represent the axis of greatest variance, while the eigenvalues associated to the eigenvectors indicate their magnitude (and thus the variance in the direction of the eigenvector). By ranking the eigenvectors in order of their eigenvalues (highest to lowest), it is possible to find the principal components in order of significance. 

The results obtained using $P_{20}$, $P_{30}$, $c$, and $w$ as input parameters are reported in Table \ref{Tab:pca_observations} and showed in Fig. \ref{fig:pca_observations}, left panel. We noticed that two components are sufficient to explain $\sim$ 90 \% of variance and that the relaxed, disturbed, and mixed systems (as defined by the visual classification, see Subsect. \ref{visual classification}), occupy different regions in the plot. For completeness, we also tested this procedure using as input parameters only $c$ and $w$ which are considered in literature among the most powerful parameters for the identification of the dynamical state of clusters. We found that the so called Principal Component 1, which is essentially a weighted combination of $c$ and $w$, is able to clearly identify the three populations of objects (Fig. \ref{fig:pca_observations}). The behaviour observed confirms that the combination of the concentration and of the centroid shift is particularly suitable for the identification of the dynamical state of clusters, as already observed by \cite{Lovisari2017}.

\begin{table}
\caption{Results obtained from the PCA.}\label{Tab:pca_observations}
\centering
\begin{tabular}{@{}ccccc@{}}
\toprule
\multicolumn{1}{l}{\begin{tabular}[c]{@{}c@{}}Input \\ parameters\end{tabular}}   & \begin{tabular}[c]{@{}c@{}}Percentage\\  (\%)\end{tabular} & \begin{tabular}[c]{@{}c@{}}Cumulative \\ percentage (\%)\end{tabular} &  &  \\ \midrule
 \multirow{3}{*}{$P_{20}$, $P_{30}$, $c$, $w$} &  67.5   & 67.5  &  &  \\ 
         &      22.6       & 90.1  &  &  \\
            &    6.9        & 97.0  &  &  \\ 
                & 3.0            & 100   &  &  \\ \midrule
 \multirow{2}{*}{$c$, $w$} &  77.7   & 77.7  &  &  \\ 
  &      22.3     & 100 &  &  \\ \bottomrule
\end{tabular}
\tablefoot{From left to right: input parameters used to realise the PCA, percentage of variance, and cumulative percentage of variance.}
\end{table}
\FloatBarrier
\section{Ring emission in simulations}\label{ring}
\begin{figure}
    \centering
    \includegraphics[scale=0.75]{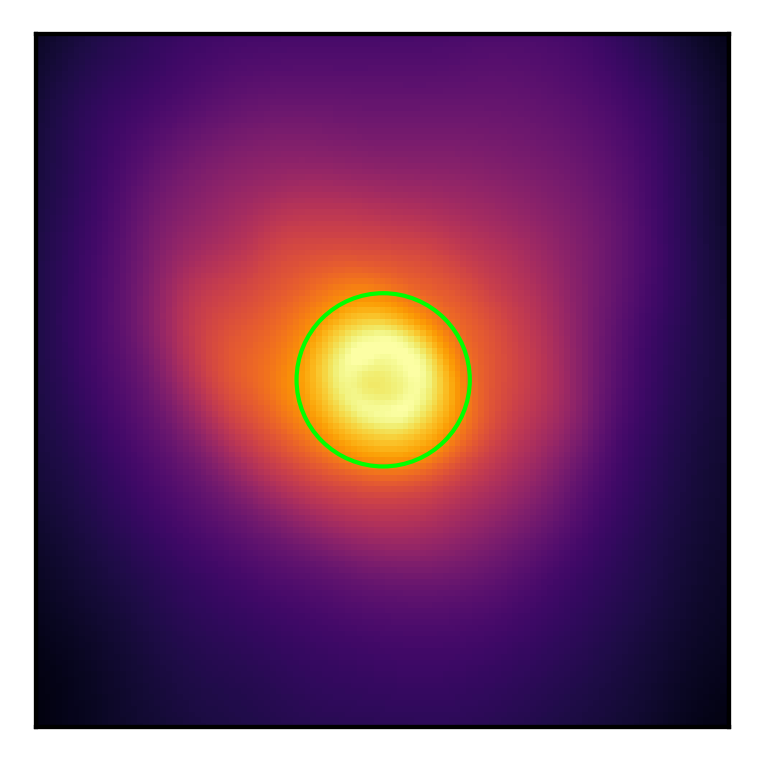}
    \caption{Example of ring emission observed in the central regions of simulated clusters. The green circle represents a circular region of radius $r$=40 kpc.}
    \label{fig:anello}
\end{figure}
In Fig. \ref{fig:anello} is shown the ring emission observed in the central regions (i.w. $r<$ 40 kpc) of the simulated clusters. As presented in Subsect. \ref{concentration_discussion}, this effect is related to the isotropic model used for the description of the AGN feedback and has an impact on the estimation of the concentrations of the simulated sample.

\end{appendix}
\end{document}